\documentclass[12pt]{article}

\usepackage{amsmath}
\usepackage{amsthm}
\usepackage{amssymb}
\usepackage{graphics}
\usepackage{verbatim}
\usepackage{bm}
\usepackage{fancyhdr}

\newtheorem{theorem}{Theorem}
\newtheorem{lemma}[theorem]{Lemma}

\newcounter{stepctr}
\newenvironment{steplist}
{\begin{list}{
Step \thestepctr:} {
\setcounter{stepctr}{0}
\usecounter{stepctr}
\setlength{\listparindent}{0em}
\setlength{\parsep}{0 em}
\setlength{\leftmargin}{3.5em}
\setlength{\itemindent}{-1em}
}}{\end{list}}

\def\defn#1{{\bf #1}}

\def\Real{{\mathbb R}}

\def\innerprod(#1,#2){{\left<#1\,,\,#2\right>}}
\def\Set#1{{\left\{#1\right\}}}

\def\Vnorm#1{\left\|#1\right\|}
\def\Snorm#1{|#1|}

\def\dual#1{{\widetilde{#1}}}
\def\man{{\cal M}}

\def\dualV{{\dual V}}

\def\calF{{\cal F}}
\def\calG{{\cal G}}
\def\calE{{\cal E}}

\def\calW{{\cal W}}
\def\calU{{\cal U}}
\def\calB{{\cal B}}
\def\calL{{\cal L}}

\def\hZ{\hat Z}
\def\epsmax{\varepsilon_{\textup{max}}}

\newcommand{\PD}{{\partial}}
\newcommand{\widedual}[1]{{\widetilde{#1}}}
\newcommand{\Vep}{{\varepsilon}}
\newcommand{\zetaVep}{{\zeta^\Vep\left(\hat{\sigma}^\Vep(t,z)\right)}}

\newcommand{\Mm}{{\frac{m_0 c^2}{\epsilon_0}}}

\title{Asymptotic Analysis of Ultra-relativistic Charge}
\author{David A. Burton, Jonathan Gratus and Robin W. Tucker\\\\Department
of Physics, Lancaster University\\
and The Cockcroft Institute}

\begin{document}
\maketitle
\begin{abstract}
This article offers a new approach for analysing the dynamic
behaviour of distributions of charged particles in an
electromagnetic field. After discussing the limitations inherent
in the Lorentz-Dirac equation for a single point particle a simple
model is proposed for a charged continuum interacting
self-consistently with the Maxwell field in vacuo. The model is
developed using intrinsic tensor field theory and exploits to the
full the symmetry and light-cone structure of Minkowski spacetime.
This permits the construction of a regular stress-energy tensor
whose vanishing divergence determines a system of non-linear
partial differential equations for the velocity and self-fields of
accelerated charge. Within this covariant framework  a particular
perturbation scheme is motivated by an exact class of solutions to
this system describing the evolution of a charged fluid under the
combined effects of both self and external electromagnetic fields.
The scheme yields an asymptotic approximation in terms of
inhomogeneous linear equations for the self-consistent Maxwell
field, charge current and time-like velocity field of the charged
fluid and is defined as an ultra-relativistic configuration. To
facilitate comparisons with existing accounts of beam dynamics an
appendix translates the tensor formulation of the perturbation
scheme into the language involving electric and magnetic fields
observed in a laboratory (inertial) frame.
\end{abstract}

\section{Introduction}
\label{ch_intro}
The intense international activity involved in probing the structure of
matter on all scales, with particle beams and radiation, owes much
to recent advances in accelerator science and technology.
Developments in the production of high power laser radiation also
offer new avenues for accelerator design and new diagnostic tools
of relevance to medical science, engineering and the
communications industry. A common theme in these developments is
the interaction between charged particles and the electromagnetic
field in domains where relativistic effects cannot be ignored.

It is remarkable that many of the challenges that must be
addressed in order to develop and control devices that accelerate
charged particles  have their origin in the interaction of
particles with their {\it own electromagnetic field}. Despite the
fact that the classical laws of electromagnetism were essentially
formulated over a century and a half ago the subject of
electromagnetic interactions with matter remains incomplete. This
incompleteness has had concomitant effects on the development of
quantum electrodynamics and renormalisation theory. At root, the
difficulties reside in the recognition that the quantum structure
of matter at some scale is beyond observation. Furthermore, the
classical description of the electron as a point particle leads
to singularities in the Maxwell self-fields that inevitably
create ambiguities in its interaction with the Maxwell field. The
general consensus is that a useful domain of validity of the
Lorentz-Dirac
equation~\cite{landau_lifshitz:1962,poisson:1999,rohrlich:1990},
describing covariantly the radiation reaction on a point electron, can be
accommodated by performing a\lq\lq reduction of order" that
effectively replaces the equation by a perturbative second order
system for the particle world line. One must then decide whether
higher order terms in this expansion should be maintained given
the neglect of terms associated with the regularisation scheme.
This approach is behind many of the successful applications of
approximate radiation reaction dynamics, despite the somewhat
delicate and unsatisfactory nature of the arguments that purport
to support this approach.

It appears that the analysis of systems involving radiating matter
has progressed by a very successful symbiosis between experimental
expediency and a variety of approximation schemes with proven
effectiveness in different domains of validity. However, as schemes
for accelerating charged particles  in these devices become more
complex and ambitious in their aims it is apparent that some
existing theoretical models are inadequate for a proper
understanding of new challenges. Such models are sometimes
enshrined in large commercial computer codes that do not   survive
scientific scrutiny or have documentation that makes contact with
established scientific literature difficult. As particle energies
increase with the use of higher intensity laser fields this issue
becomes critical and more reliable methods for accommodating
radiation reaction must be found.

A mathematically coherent formulation of a closed system of
partial differential equations describing the relativistic
behaviour of charged matter with electromagnetic fields is
inevitably non-linear and, in general, exact solutions satisfying
causal boundary conditions are intractable. The fact that the
effects of (coherent and incoherent)  radiation on
micrometre-sized
charged bunches in a host of newly proposed advanced devices is  a
significant barrier to development indicates that existing
approaches used to model such effects are inadequate and that a
new look at the whole problem is timely.

Many of the radiation problems alluded to above can be
circumvented by working {\it entirely in the language of fields}
rather than both particles and fields. This makes it possible to
use intrinsic tensor field theory  and exploit to the full the
power of differential geometry in a relativistic spacetime
framework.  The geometric field formulation offers a number of
powerful computational advantages over existing formulations of
beam dynamics. These include the maintenance of relativistic
covariance (local Lorentz transformation between local
(accelerating) frames are unnecessary), the use of curvilinear
coordinates adapted to the geometry of particular problems
(thereby facilitating imposition of boundary conditions),
exploitation of available symmetries and the formulation of
coordinate-independent approximation schemes.

In this article a new approach is offered for analysing the
behaviour of distributions of charged particles in a coupled
electromagnetic field environment. Guided by a simple model, an
approximation scheme designed for charged distributions containing
\emph{ultra-relativistic} particles\footnote{A novel definition of
an ``ultra-relativistic vector field'' relying entirely on the
light-cone structure of spacetime is given in Appendix
\ref{appendix:ultrarel_light-like_defs}.} is explored. The exact
equations of motion are derived from the vanishing divergence of a
relativistic stress-energy tensor for matter and radiation.
Before presenting the approximation scheme in detail we recall the
genesis of the radiation reaction for a {\em point} charge,
mention its shortcomings and argue that these may be overcome by
treating a large collection of charged particles as a {\em
continuum}.


\subsection{Equations of motion for radiating charges}
\label{ch rad reaction}
\def\q{q_0}
\def\kk{{\frac{\q^2}{4\pi\epsilon^2_0}}}
\def\kkk{{\frac{\q^2}{4\pi\epsilon_0}}}
\def\gg{{\cal G}_X}
\def\P{{\cal P}}
\def\nn{{\cal N}}

\def\BE#1{\begin{equation} #1 \end{equation} }

In this paper the language of differential geometry is employed
since it offers the most succinct mathematical framework for
electrodynamics. Thus, all fields will be regarded as sections of
tensor bundles over appropriate domains ${\cal M}$ of Minkowski
spacetime endowed with a
fixed metric $g$ and a torsion-free metric compatible connection
$\nabla$. Sections of the tangent bundle over ${\cal M}$ will be denoted $\Gamma
T{\cal M}$ while sections of the bundle of exterior $p$-forms will
be denoted $\Gamma \Lambda^p{\cal M}$. For any vector field $X$
denote by $\widetilde X$ the associated $1$-form defined by
$\widetilde X=g(X,-)$. The operator $d$ will denote the exterior
derivative and $i_X$ the contraction operator with respect to $X$.
To facilitate comparisons with existing accounts of beam dynamics
Appendix \ref{appendix:Gibbs_3-vector_version} translates the tensor
formulation of the perturbation
scheme into the $3$-vector language involving electric and magnetic fields
observed in a laboratory (inertial) frame.

Due to its high symmetry,
Minkowski spacetime admits a class of global charts that play a
fundamental role in the following. A generic system of coordinates
in one of these charts will be denoted $\{y^\mu\}\in\Real^4$ where
$\mu$ ranges from $0$ to $3$. In these coordinates the metric
tensor takes the form \BE{g=-d\,y^0\otimes d\,y^0 +d\,y^1\otimes
d\,y^1 +d\,y^2\otimes d\,y^2 +d\,y^3\otimes d\,y^3.} With $y^0>0$
the field $\frac{\PD}{\PD y^0}$ is defined to be future-pointing.
When convenient $y^0=ct, y^1=x,y^2=y, y^3=z$ is written and the
vacuum speed of light $c$ is set to $1$. The significance of this class of
coordinates is that it offers a basis of symmetry generators
$\{K_\mu=\frac{\PD}{\PD y^\mu}\}$:\BE{{\cal L}_{K_\mu} g =0}
where ${\cal L}_X$ is the Lie derivative with respect to $X$.
Such Killing vectors will be used to define energy-momentum
densities and power-momentum fluxes associated with different
field configurations on spacetime.

If $\calU$ is a domain of spacetime  with boundary \BE{\partial
\calU= \Sigma_1+\Sigma_2 + \Pi} for space-like hypersurfaces
$\Sigma_1,\Sigma_2$ and ${\cal J}$ a closed regular 3-form on $\calU$ (i.e.
$d{\cal J}=0$) then \BE{\int_{\partial\calU} {\cal J}=
\int_{\calU} d\,{\cal J}=0.}
 Thus
\BE{\int_{\Sigma_1}{\cal J} = \int_{-\Sigma_2}{\cal J} - \int_\Pi
{\cal J}.} Such closed 3-forms imply conservation laws.

In a vacuum the electric and magnetic fields are encoded into the
$2$-form $\calF$ and the electric charge density and current are
described by a source current 3-form $j$.
The Maxwell field system on spacetime is
\BE{d\,\calF=0\label{maxeq1}} and
\BE{d\,\star\calG=-j\label{maxeq2}}
where $\star$ is the Hodge map associated
with the spacetime metric tensor $g$ and $\calG=\epsilon_0\calF$
where $\epsilon_0$ is the permittivity of the vacuum.

For any vector field $W$ on spacetime and any Maxwell solution
$\calF$ define the ``electromagnetic drive'' 3-form \BE{
\tau^{\text{(EM)}}_W=\frac{1}{2}\{i_W\calF \wedge \star\calF -
i_W\star\calF \wedge \calF\}\label{EM_stress_forms}.} If $W$ is a
Killing vector field then $\tau^{\text{(EM)}}_W$ is called a ``Killing
current'' and \BE{
d\,\tau^{\text{(EM)}}_W=\frac{1}{\epsilon_0}i_W\calF\wedge j} follows
from (\ref{maxeq1}), (\ref{maxeq2}) and (\ref{EM_stress_forms}).
For each Killing vector field these equations establish a \lq\lq
local conservation equation\rq\rq ( $d\tau_W=0 $) in a source free
region $(j=0)$.

If $U$ is any time-like vector field (with $g(U,U)=-c^2$) one may write
uniquely \BE{\calF=\tilde e \wedge \tilde U + \star(\tilde b\wedge
\tilde U)} where $g(e,U) = g(b,U) = 0$
and \BE{\frac{1}{c^2}\tau^{\text{(EM)}}_U=-\tilde e\wedge \tilde b \wedge \tilde U -
\frac{1}{2} \{ g(e,e) + g(b,b)\} i_U(\star 1).}

Poynting\footnote{In a pre-relativistic context.} related the form  $\tilde e\wedge \tilde b$, in a source free region,
to the local field energy transmitted normally across unit area
per second (spatial field energy current or field power) and $\frac{1}{2} \{ g(e,e) +
g(b,b)\}$ to the local electromagnetic field energy density.

More precisely $\frac{1}{c}\int_\Sigma \tau^{\text{(EM)}}_U$ is the field energy
associated with the space-like 3-chain $\Sigma$ and
$\frac{1}{c^2}\int_{S}\,i_U\tau^{\text{(EM)}}_U$ is the power flux across an
oriented space-like 2-chain $S$.

If $X$ is a space-like Killing vector generating space-like
translations along open integral curves then with the split:
\BE{\tau^{\text{(EM)}}_X=\mu_X \wedge \tilde U + \gg,} where $i_U\mu_X=0$ and
$i_U\gg=0$, the Maxwell stress
2-form $\mu_X$   may be used to identify mechanical forces
produced by a flow of field momentum  with density 3-form $\gg$.
In any local frame $\{X_a\in \Gamma T{\cal M }\}$ with dual
coframe $\{e^b\in \Gamma \Lambda^1{\cal M }\}$ the 16 functions
$T^{\text{(EM)}}_{ab}=i_{X_b} \star \tau^{\text{(EM)}}_{X_a}$, where the frame
indices $a,b=0,1,2,3$, may be used to
construct the second-rank stress-energy tensor
$$T^{\text{(EM)}}=T^{\text{(EM)}}_{ab} \,e^a\otimes e^b.$$

A standard application of these general notions is the derivation
of the equation of motion of a charged fluid in an external
electromagnetic field from the total stress-energy  of the fluid
and electromagnetic field. A thermodynamically inert (cold) fluid
can be modelled with the stress-energy tensor
\BE{T^{\text{(f)}}=\frac{m_0}{c\epsilon_0} \,\nn \widetilde V \otimes
\widetilde V } where $\nn$ is a scalar number density field, $m_0$
some constant with the dimensions of mass, $V$ the unit
time-like 4-velocity field of the fluid and $g(V,V)=-1$. Such a
stress-energy tensor gives rise to a set of Killing currents\BE{
\tau^{\text{(f)}}_{K_\mu}=  \frac{m_0}{c\epsilon_0} g(V,K_{\mu}) \star
(\nn \widetilde V) } which are added to $ \tau^{\text{(EM)}}_{K_\mu}$ to
yield the total set of Killing currents for the interacting
system. If one assumes that the electric current $3$-form is
$j=q_0\nn \star \widetilde V$ for some electric charge constant
$q_0$ and that $\nn$ is \emph{regular} then the conservation laws
\BE{d\,j=0}
 \BE{ d(\tau^{\text{(EM)}}_{K_\mu}  +
\tau^{\text{(f)}}_{K_\mu})=0} yield the field equation of motion
\BE{\nabla_V \widetilde V= \frac{q_0}{m_0 c^2} i_V \calF.
\label{max_lor}} This equation must be solved consistently with
the Maxwell equations (\ref{maxeq1}), (\ref{maxeq2}) to determine
$V$, $\nn$ and $\calF$ for prescribed initial and boundary conditions.

If the source of $\calF$ contains a {\em point charge} in an
external field this approach must take into account that part of
the electromagnetic field becomes singular on the world-line of
the particle so the domain $\calU$ above must exclude this line.
Furthermore, one needs to postulate a stress-energy tensor for a
stable point particle.  Without further information application of
Stokes' theorem then leaves an unknown contribution to the
conservation law. However, experiment indicates that accelerating
electrons (thought to be point particles) experience a reaction to
their emitted  radiation  and that the power radiated is in good
agreement with the covariant Larmor formula \cite{jackson:1998}.
A number of methods have been devised in an attempt to accommodate
this observed radiation reaction by adopting a regularisation
procedure that permits  the use of Stokes' theorem in a spacetime
region that includes the particle world-line. Inevitably this
requires some assumption of how the singular Coulombic  stresses
are compensated by stresses that are not of  electromagnetic
origin.
 Early approaches invoked delicate limiting processes and the use
of both advanced as well as retarded solutions to Maxwell's
equations {\cite{dirac:1938, rohrlich:1990}}. Dirac \cite{dirac:1938}
offered one of the simplest covariant regularisation schemes and
the resulting equation of motion is known as the Lorentz-Dirac
equation. The use of both advanced and retarded solutions is
however unnecessary \cite{teitelboim:1980, flanagan_wald:1996}.

To highlight the assumptions in the single particle approach to
radiation reaction an outline is given of how the Lorentz-Dirac equation can be
derived using a coordinate system adapted to the light-cone
structure of spacetime and the time-like world-line of a particle
on a parametrised history given in the above chart as
$y^\mu=\xi^\mu(u)$. For each value of $u$ the unit tangent to the
world-line is the 4-velocity \BE{V=V^\mu(u)\frac{\PD}{\PD y^\mu}}
where $V^\mu(u)=\frac{d\xi^\mu(u)}{du}$.

Assign to any event in spacetime not on the particle world-line,
whose backward light-cone intersects this world-line at the event
with parameter $u$, the set of coordinates $\{ u,r,\theta,\phi\}$
with $0\le r\le \infty, 0<\theta\le \pi, 0< \phi \le 2\pi$ using
the transformation
\BE{y^\mu=\xi^\mu(u)+\frac{r}{{\P}(u,\theta,\phi)}L^\mu(\theta,\phi),}
where for any $y^0>0:$ \BE{L=\frac{\PD}{\PD y^0} + \sum_{k=1}^3
n^k(\theta,\phi)\frac{\PD}{\PD y^k} =L^\mu\,\frac{\PD}{\PD y^\mu}
} \BE{ \sum_{k=1}^3 (n^k(\theta,\phi))^2 =1}
 \BE{\P=-g(V,L), \qquad g(V,V)=-1, \qquad g(L,L)=0.}
In these coordinates the particle world-line is the curve where\footnote{For
$\P>0$ with $V$ time-like and $\frac{\PD}{\PD y^0}$ future-pointing the
set $r\ge 0$ lies on the forward light-cone of the event $\xi^\mu(u)$.}
$r=0$, $(\theta,\phi)$ parametrise the unit $2$-sphere $S^2$ where
$\{n^k(\theta,\phi)\}$ are direction cosines and the metric tensor can
be written: \BE{g=-e^0\otimes e^0
+e^1\otimes e^1 +e^2\otimes e^2 +e^3\otimes e^3} where \BE{ e^0=
\calB\,d\,u -\frac{1}{\calB}\,d\,r} \BE{e^1=\frac{d\,r}{\calB}}
\BE{e^2=\frac{r}{\P}d\,\theta}
\BE{e^3=\frac{r}{\P}\sin\theta\, d\,\phi  }
\BE{\calB=\sqrt{1+2r\frac{\dot \P}{\P}}.} Here, and in the following,
denote $\PD_u f$ by $\dot f$ for any scalar field $f$ and let $\dot
{\cal T}= {\cal L}_{\frac{\PD}{\PD u} }{\cal T}$ for any tensor
field ${\cal T}$.


Suppose the total system (electromagnetic field and charge) has a stress-energy
tensor giving rise to a \emph{regular} Killing current $\tau_K$ such
that \BE{d\,\tau_K=0} in a domain $\calU$ bounded by the light-like
3-chains $\Sigma(u=u_0 + \Delta u)$, $\Sigma(u=u_0)$ and the time-like 3-chain $
\Sigma(r=r_0)$ for arbitrary positive constants $u_0,\Delta
u,r_0$. Then \BE{\int_{\Sigma(u=u_0+\Delta u)}\tau_K -
\int_{\Sigma(u=u_0)}\tau_K + \int_{\Sigma(r=r_0)}\tau_K=0.}

In the limit $\Delta u \to 0$ this may be written \BE{\dot
P^{\text{(B)}}_K(u_0,r_0)\, d\,u + \dot P^{\text{(C)}}_K(u_0,r_0) \,
d\,u =0} where \BE{\dot P^{\text{(B)}}_K(u_0,r_0)= \int_0^{r_0}\int_{S^2}
\dot \tau_K(u_0,r,\theta,\phi) } \BE{ \dot P^{\text{(C)}}_K(u_0,r_0)=
\int_{S^2} i_{\PD_u}\tau_K(u_0,r_0,\theta,\phi)\label{P_K_C_dot} }

With $K\in\{ \frac{\PD}{\PD y^\mu} \} $ and $\nabla \frac{\PD}{\PD
y^\mu} =0$ one has a similar balance of rates for each
translational Killing vector in the basis. In terms of the
``body'' rate  \BE{ \dot P^{\text{(B)}}\equiv \dot
P^{\text{(B)}}_{\frac{\PD}{\PD y^\mu}}\,d\,y^\mu } and the \lq\lq
contact" rate \BE{ \dot P^{\text{(C)}}\equiv\dot P^{\text{(C)}}_{\frac{\PD}{\PD
y^\mu}}\,d\,y^\mu } there is a balance of $1$-forms: \BE{\dot
P^{\text{(B)}}(u_0,r_0)+\dot P^{\text{(C)}}(u_0,r_0)=0\label{balance}.}

Next assume that the contact rate includes a part from the Killing
currents $\tau^{\text{(EM)}}_K$, i.e. \BE{\dot P^{\text{(C)}}=\dot
P^{\text{(C)}}_{\text{MECH}}+\dot P^{\text{(C)}}_{\text{EM}}} and
furthermore that this part
can be calculated from the \emph{retarded} Li\'enard-Wiechert
solution ${\calF}$ to Maxwell's equations for an arbitrarily
moving point charge $\q$ in no external electromagnetic field. In
these coordinates the solution is ${\calF}=dA$ where the 1-form
\BE{ A=\frac{\q}{4\pi\epsilon_0} \frac{\widetilde V(u)}{r}.}
Integration over the $2$-chain $S^2$ with $r=r_0$ yields for
(\ref{P_K_C_dot}) : \BE{ \dot P^{\text{(C)}}_{\text{EM}}= \kk\left\{
\frac{2}{3}\widetilde{\cal A}({\cal A})\,\widetilde V +
\frac{\widetilde{\cal A}}{2r_0} \right\}\biggm|_{u=u_0}} where
${\cal A}=\dot V^\mu(u)\frac{\PD}{\PD y^\mu}$ is the acceleration
field. Clearly this 1-form is singular on the world-line where
$r_0=0$. The first term however correctly accounts for the
observed Larmor radiation rate of energy-momentum from an
accelerating charge and is independent of $r_0$. To proceed one
must cancel the singular rate in $\dot P^{\text{(C)}}_{\text{EM}}$ from singular
terms  in the remaining rates in such a way that in the limit
$r_0\to 0$ the resulting system of ordinary differential equations
for arbitrary $u_0$ makes both mathematical and physical sense.
One approach is to suppose that the remaining rates are determined
in terms of scalar fields $\alpha(u,r), \beta(u,r)$ and the
vectors $V,{\cal A}$ such that \BE{ P^{\text{(B)}}(u,r_0)+
P^{\text{(C)}}_{\text{MECH}}(u,r_0)=\alpha(u,r_0)\widetilde V(u,r_0)
 + \beta(u,r_0) \widetilde{\cal A}(u,r_0).}
Inserting this in (\ref{balance}) and applying the projection
operator $\Pi_V=1+\widetilde V \wedge i_V$   one finds
\BE{\beta=-\frac{2}{3}\kk + \frac{\dot\alpha}{\widetilde{\cal
A}({\cal A})}} With this value (\ref{balance}) yields
\BE{\widetilde{\cal A} \left[\alpha -\frac{d}{du}\left(\frac{\dot
 \alpha}{\widetilde{\cal A}({\cal A})}\right)
 +\kk\frac{1}{2r_0}\right ] =-\widetilde
 V\left[\dot \alpha + \frac{2}{3}\kk \widetilde{\cal A}({\cal A})\right] +
 \left[\frac{2}{3}\kk
 + \frac{\dot \alpha}{\widetilde{\cal A}({\cal A})}\right ]\widetilde{\dot{\cal
 A}}}
where ${\dot{\cal A}}=\ddot V^\mu(u)\frac{\PD}{\PD y^\mu}$. One
way to simplify this and cancel the exposed electromagnetic
singularity is to further assume that
 $\alpha(u,r_0)=\Mm-\kk\frac{1}{2r_0}$ for some {\em constant}
$m_0$ so that in the limit $r_0\to 0$ one has: \BE{m_0
c^2\widetilde{\cal A}=  \frac{2}{3}\kkk\,
\Pi_V\widetilde{\dot{\cal A}}\label{raw}} or \BE{m_0
c^2\widetilde{\cal A}= - \frac{2}{3}\kkk\, i_V(\widetilde V\wedge
\widetilde{\dot{\cal A}}).} This is a manifestly covariant
equation of motion even though the velocity and acceleration are specified in
a basis of Killing vector fields. Since it is tensorial it is
independent of basis and in terms of covariant differentiation
along the world-line it may be written \BE{m_0 c^2\widetilde{\cal
A}= - \frac{2}{3}\kkk\, i_V(\widetilde V\wedge
{\nabla_V\widetilde{\cal A}}).\label{coveqn}} This is a system of
third order differential equations for the worldline $\xi^\mu(u)$.
Regarded as an initial value problem it requires unfamiliar
initial data ($\xi^\mu(0), \dot\xi^\mu(0),\ddot\xi^\mu(0)$) and
solutions exist corresponding to self acceleration which must be
regarded as unphysical.

In the presence of an external Maxwell field ${\calF}_{\text{ext}}$ one
must confine the motion of the particle to a domain $\calU$ that
excludes the sources of ${\calF}_{\text{ext}}$. In this situation one
expects that the equation of motion will acquire a contribution
from the Lorentz force $\q i_V {\calF}_{\text{ext}}$: \BE{m_0
c^2\widetilde{\cal A}=\q i_V {\calF}_{\text{ext}}+ \frac{2}{3}\kkk\,
\Pi_V\widetilde{\dot{\cal A}}\label{lordirac}} so that $m_0$ is
identified with the rest mass of the point particle.

Although solutions to this system that self-accelerate can be
eliminated by demanding contrived data at different points along
the world-line there remain solutions  that pre-accelerate in
situations where the external field is piecewise defined in
spacetime {\cite{rohrlich:1990}}. Although the duration of
pre-acceleration is probably classically unobservable for most
electrons this general feature suggests that not all of the
assumptions above are acceptable.

One resolution of these difficulties {\cite{landau_lifshitz:1962}}
is to assume that the right hand side of (\ref{lordirac}) should
be expanded as a series in $\q$ with leading term for $\tilde
{\cal A}$ given by $\frac{\q}{mc^2} i_V {\calF}_{\text{ext}}$. Then the
above assumptions are held to be accurate only to some order in
$\q$ and (\ref{lordirac}) is understood as: \BE{\widetilde{\cal
A}= \frac{\q}{m_0 c^2} i_V \calF_{\text{ext}}- \frac{2}{3m_0 c^2}\kkk\,
i_V(\widetilde V\wedge {\nabla_V\widetilde{\cal A}_{\text{ext}}})+\ldots
\label{newcoveqn}} where $\widetilde {\cal A}_{\text{ext}}=\frac{\q}{m_0 c^2}
i_V \calF_{\text{ext}} $. The system is now manifestly a second  order
system of evolution equations. Although this offers a workable
scheme it is unclear what its limitations are in different types
of external field. Furthermore in situations where one has to
contemplate the radiation from a large number of accelerating
high-energy particles in close proximity the neglect of higher
order terms in the expansion may be suspect.

Given the complexities and reservations associated with
(\ref{newcoveqn}) compared with the model leading to
(\ref{max_lor})  the latter is adopted in this article as a
description of a collection of potentially radiating particles in
the presence of an external electromagnetic field. Thus, our
fundamental set of equations is
\begin{align}
&d \calF = 0\,,
\label{np_maxwell_V_d_F_unscaled}
\\
&d \star \calF = - \frac{q_0}{\epsilon_0}{\cal N}\star\dualV \,,
\label{np_maxwell_V_d_star_F_unscaled}
\\
&\nabla_V \dual{V} = \; \frac{q_0}{m_0 c^2}{i_{V} \calF}\,,
\label{np_maxwell_V_lorentz_unscaled}
\\
&V\cdot V=-1
\label{np_maxwell_V_V_dot_V_unscaled}
\end{align}
where $X\cdot Y \equiv g(X,Y)$ for all vector fields $X, Y\in\Gamma T\man$.

To simplify the subsequent analysis the system
(\ref{np_maxwell_V_d_F_unscaled}-\ref{np_maxwell_V_V_dot_V_unscaled})
is recast using the $2$-form $F=\frac{\q}{m_0 c^2}\calF$ and the scalar
field $\rho=\frac{\q^2}{\epsilon_0 m_0 c^2}{\cal N}$ where $\q
{\cal N}$ is the proper charge density since ${\cal N}$ is the
proper number density. This yields the system of field equations
\begin{align}
&d F = 0\,,
\label{np_maxwell_V_d_F}
\\
&d \star F = - \rho \star \dualV \,,
\label{np_maxwell_V_d_star_F}
\\
&\nabla_V \dual{V} = \; {i_{V} F}\,,
\label{np_maxwell_V_lorentz}
\\
&V\cdot V=-1
\label{np_maxwell_V_V_dot_V}
\end{align}
for the triple $(V,\rho,F)$. Equation (\ref{np_maxwell_V_d_star_F})
leads immediately to the integrability condition (conservation of
electric charge)
\begin{align}
d\star (\rho\dual{V})=0.
\label{np_maxwell_d_rho_star_V}
\end{align}
The scalar field  $\rho$ will be referred to as the \emph{reduced}
proper charge density. Since ${\cal N}$ is a number density,
${\cal N}\ge 0$ and so $\rho\ge 0$ for $q_0>0$ and $q_0<0$.
\newcommand{\Fieldsysrefs}{(\ref{np_maxwell_V_d_F}-\ref{np_maxwell_V_V_dot_V})
}
\newcommand{\Fieldsysintegrefs}{(\ref{np_maxwell_V_d_F}-\ref{np_maxwell_d_rho_star_V})
}

\section{Symmetric solutions to the coupled field system}
\label{ch_maxlor11}
Although the field system \Fieldsysrefs
appears considerably less complicated than a
large number of ordinary differential equations for a collection of
accelerating charges, it is non-linear in $(V,\rho,F)$ and
so obtaining exact solutions is difficult except in
certain simple circumstances. One method of obtaining solutions
in more general situations is to employ a perturbation scheme based on
properties of a particular class of exact solutions.

In this section  a family of \emph{exact} dynamical solutions with
symmetries generated by $\frac{\PD}{\PD y^1}$ and $\frac{\PD}{\PD
y^2}$ will be described. Such highly symmetric solutions can be
interpreted as accelerating ``walls of charge'' moving in vacuo
under the influence of their self-fields (their ``space-charge''
fields) and an externally applied field.
 The coordinate $y^0$ will be identified with time $t$ in an
 inertial laboratory frame (the speed of light
$c=1$) and $x\equiv y^1$, $y\equiv y^2$, and $z\equiv y^3$. The solutions
will be independent of $y^1$ and $y^2$ i.e.
$${\cal L}_{\frac{\PD}{\PD y^1}} F=  {\cal L}_{\frac{\PD}{\PD y^2}} F =   0,$$
$${\cal L}_{\frac{\PD}{\PD y^1}} V=  {\cal L}_{\frac{\PD}{\PD y^2}} V=    0.$$
Thus one may reduce the problem to a field theory on a
2-dimensional Lorentzian spacetime with global coordinates
$y^0,y^3$.  The field system \Fieldsysrefs is solved exactly
using a co-moving coordinate system $(\tau,\sigma)$ adapted to the
charged continuum. However, expressing the solutions in terms of
laboratory coordinates $(t,z)$ requires the inverse of the mapping
$(\tau,\sigma)\rightarrow (t,z)$, which is generally difficult to
obtain in closed form.
To progress, a running parameter $\Vep>0$ is introduced into the
mapping $(\tau,\sigma)\rightarrow (t,z)$ and a perturbation scheme in
$\Vep$ is established.
This approach facilitates an order-by-order
construction of the inverse of the mapping $(\tau,\sigma)\rightarrow
(t,z)$ and thereby leads to $1$-parameter families $(V^\Vep,\rho^\Vep,F^\Vep)$ of
solutions in $\Vep$. It will be shown that
\begin{equation*}
F^\Vep = \sum\limits^\infty_{n=-1} \Vep^nF_n,\quad
V^\Vep = \sum\limits^\infty_{n=-1} \Vep^nV_n,\quad
\rho^\Vep = \sum\limits^\infty_{n=1} \Vep^n\rho_n
\end{equation*}
over some range of $\Vep$
where the coefficients $F_n$, $V_n$ and $\rho_n$ are 2-forms, vector
fields and scalar fields respectively.

Exact solutions to the system of equations \Fieldsysrefs are obtained using the ans\"atz
\begin{align*}
&F = \calE(t,z)\, dt\wedge dz,\\
&V = \frac{1}{\sqrt{1-\mu^2(t,z)}}\left(\PD_t + \mu(t,z)\PD_z\right)
\end{align*}
where $\mu$ is the magnitude of the Newtonian velocity field of
the charged continuum measured by the inertial (laboratory)
observer $\PD_t$. Equations (\ref{np_maxwell_V_d_F}) and
(\ref{np_maxwell_V_V_dot_V}) are manifestly satisfied while
(\ref{np_maxwell_V_d_star_F}) and (\ref{np_maxwell_V_lorentz})
lead to
\begin{align}
\label{maxlor11_d_E}
&d\calE = \rho\# \dual{V},\\
\label{maxlor11_V_lorentz}
&\nabla_V \dual{V} = \calE \#\dual{V}
\end{align}
where $\#$ is the Hodge map associated with the volume $2$-form
$\# 1 \equiv dt\wedge dz$. Since $i_V\# \alpha = \#(\alpha\wedge
\dual{V})$ for all forms $\alpha$ independent of $dx$ and
$dy$, the action of $i_V$ on (\ref{maxlor11_d_E}) yields
\begin{equation}
\label{maxlor11_E_const_along_V}
V\calE = 0
\end{equation}
i.e. $\cal{E}$ is constant along the integral curves of $V$. Let
$C_\sigma$ be a $1$-parameter family of proper-time-parametrised integral
curves of $V$ where each value of $\sigma$ corresponds to an integral
curve of $V$:
\begin{equation*}
\begin{split}
C_\sigma : \mathbb{R} &\rightarrow \man,\\
\tau &\rightarrow \left( t=\hat{t}(\tau,\sigma),z=\hat{z}(\tau,\sigma)\right)
\end{split}
\end{equation*}
and
\begin{equation*}
V = C_{\sigma*}\PD_\tau = \frac{\PD\hat{t}}{\PD\tau}\PD_t +
\frac{\PD \hat{z}}{\PD\tau}\PD_z
\end{equation*}
where $C_{\sigma*}$ is the push-forward (tangent)
map associated with $C_\sigma$. Equation (\ref{maxlor11_E_const_along_V}) is
\begin{equation*}
\begin{split}
(C_{\sigma*}\PD_\tau)\calE &= \PD_\tau(C_\sigma^*\calE)\\
&= 0
\end{split}
\end{equation*}
where $C^*_\sigma$ is the pull-back
map associated with $C_\sigma$. Hence $C_\sigma^*\calE$ depends only
on $\sigma$ and can be expressed using the function
$\zeta:\mathbb{R}\rightarrow\mathbb{R}$ where
\begin{equation}
\label{maxlor11_zeta_sigma}
\begin{split}
\zeta(\sigma) &\equiv C_\sigma^*\calE\\
&= \calE\left(\hat{t}(\tau,\sigma),\hat{z}(\tau,\sigma)\right).
\end{split}
\end{equation}
Using $\nabla\PD_t=\nabla\PD_z=0$ the $(t,z)$ components of
(\ref{maxlor11_V_lorentz}) are
\begin{align}
\notag
&\frac{\PD^2\hat{t}}{\PD\tau^2}(\tau,\sigma) = \zeta(\sigma)\frac{\PD\hat{z}}{\PD\tau}(\tau,\sigma),\\
\label{maxlor11_ODES_for_t_and_sigma}
&\frac{\PD^2\hat{z}}{\PD\tau^2}(\tau,\sigma) = \zeta(\sigma)\frac{\PD\hat{t}}{\PD\tau}(\tau,\sigma)
\end{align}
whose particular solution satisfying the initial conditions
\begin{align*}
&\hat{t}(0,\sigma) = 0,\,\,
\hat{z}(0,\sigma) = \sigma,\\
&\frac{\PD\hat{t}}{\PD\tau}(0,\sigma) = 1,\,\,
\frac{\PD\hat{z}}{\PD\tau}(0,\sigma) = 0,
\end{align*}
at $\tau = 0$ is
\begin{align}
\label{maxlor11_t_tau_sigma}
&t = \hat{t}(\tau,\sigma) = \frac{1}{\zeta(\sigma)}
\sinh(\zeta(\sigma)\tau)\,,
\\
\label{maxlor11_z_tau_sigma}
&z = \hat{z}(\tau,\sigma) = \frac{1}{\zeta(\sigma)}
\left[\cosh(\zeta(\sigma)\tau) - 1\right] + \sigma
\end{align}
where the function $\zeta:\mathbb{R}\rightarrow\mathbb{R}$
must be supplied as data (i.e. $\zeta$ is an ingredient in the initial conditions).
Note that the charged continuum is taken to be initially at rest in
the laboratory, i.e.
\begin{equation*}
V\bigm|_{\tau=0} = (C_{\sigma*} \PD_\tau)\bigm|_{\tau=0} = \PD_t.
\end{equation*}
More general initial conditions could have been used, but the above
scenario is sufficient.

After $\zeta$ has been
specified, expressions for $V$ and $F$ in the $(t,z)$
coordinate system are obtained by inverting (\ref{maxlor11_t_tau_sigma}) and
(\ref{maxlor11_z_tau_sigma}) to give $(\tau,\sigma)$ in terms of
$(t,z)$:
\begin{align*}
&\tau = \hat{\tau}(t,z),\\
&\sigma = \hat{\sigma}(t,z)
\end{align*}
and using (\ref{maxlor11_zeta_sigma}) it follows that
\begin{equation*}
\calE(t,z) = \zeta\left(\hat{\sigma}(t,z)\right).
\end{equation*}

An initial condition on $\calE$ is expressed using $\zeta$.
Equation (\ref{maxlor11_t_tau_sigma}) is used to show that the local
hypersurfaces $t=0$ and $\tau=0$ are equal and so
(\ref{maxlor11_z_tau_sigma}) yields
\begin{equation*}
\hat{\sigma}(0,z) = z.
\end{equation*}
Thus, (\ref{maxlor11_zeta_sigma}) leads to
\begin{equation*}
\begin{split}
\calE(0,z) &= \zeta\left(\hat{\sigma}(0,z)\right)\\
&= \zeta(z)
\end{split}
\end{equation*}
on the space-like hypersurface $t=0$.

Not all choices for $\zeta$ are admissible because, as noted in the
derivation of \Fieldsysrefs\hspace{-0.4em}, the
reduced proper charge density $\rho$ is positive:
\begin{equation}
\label{maxlor11_pos_def_rho}
\rho = -i_V(\# d \calE) \ge 0
\end{equation}
using (\ref{maxlor11_d_E}) and (\ref{np_maxwell_V_V_dot_V}).
At $t = \tau = 0$
\begin{equation}
\label{maxlor11_initial_rho}
\begin{split}
\rho(0,z) &= -i_{\PD_t}\#( \PD_t\calE(0,z)dt + \PD_z\calE(0,z)dz)\\
&= \PD_z\calE(0,z)\\
&= \zeta^\prime(z)
\end{split}
\end{equation}
where $\# dt=-dz$ and $\# dz=-dt$ have been used and
$\zeta^\prime(z) \equiv \frac{d\zeta}{dz}(z)$. Hence, employing
(\ref{maxlor11_pos_def_rho})
\begin{equation*}
\zeta^\prime(z) \ge 0
\end{equation*}
i.e. $\zeta$ is a monotonically increasing function.

The coordinate chart $(\tau,\sigma)$ has important physical
significance because it labels the streamlines of the charged
continuum (the integral curves of $V$). Physically,  the
coordinate transformation $(\tau,\sigma)\rightarrow(t,z)$  is
expected to be valid over the entire $(\tau,\sigma)$ plane. This will now  be shown.

The determinant $\Delta$ of the Jacobian matrix of the coordinate transformation $(\tau,\sigma)\rightarrow(t,z)$ is
\begin{equation}
\label{maxlor11_det_jac_tau-sigma_t-z}
\begin{split}
\Delta &\equiv \text{det}\left[
\begin{array}{cc}
\frac{\PD\hat{t}}{\PD\tau} & \frac{\PD\hat{z}}{\PD\tau}\\
\frac{\PD\hat{t}}{\PD\sigma} & \frac{\PD\hat{z}}{\PD\sigma}
\end{array}
\right]\\
&= -\frac{\zeta^\prime(\sigma)}{\zeta(\sigma)^2}
+ \cosh(\zeta(\sigma)\tau)
\left(1+\frac{\zeta^\prime(\sigma)}{\zeta(\sigma)^2}\right).
\end{split}
\end{equation}
Since $\zeta^\prime(\sigma)\ge 0$ and $\cosh(\zeta(\sigma)\tau)\ge 1$ it follows that
\begin{equation*}
\Delta \ge 1.
\end{equation*}
Monotonicity of $\zeta$ implies that $\zeta$ must vanish somewhere. Let
$\sigma_0\in\Real$ be the point where $\zeta(\sigma_0)=0$ and note
\begin{equation*}
\underset{\sigma\rightarrow\sigma_0}{\lim}\Delta
= 1 + \frac{1}{2}\zeta^\prime(\sigma_0)\tau^2
\end{equation*}
using (\ref{maxlor11_det_jac_tau-sigma_t-z}). Therefore, $\Delta$ is
finite for all $(\tau,\sigma)\in\Real^2$ and the coordinate
transformation $(\tau,\sigma)\rightarrow (t,z)$ is valid over the
entire $(\tau,\sigma)$ plane.

 The implicit equation
\begin{equation}
\label{maxlor11_sigma_imp}
\sigma = z -
\frac{1}{\zeta(\sigma)}\left(\sqrt{1+\left[\zeta(\sigma)\right]^2
t^2}-1\right)
\end{equation}
for $\sigma = \hat{\sigma}(t,z)$ arises by combining
(\ref{maxlor11_t_tau_sigma}) with (\ref{maxlor11_z_tau_sigma}).

The initial electric field
$\calE(0,z)=\zeta(z)=\zeta_\textup{ext}+\zeta_\textup{self}(z)$ where
$\zeta_\textup{ext}$ models an externally applied field and
$\zeta_\textup{self}$ is the self-field of the charged continuum.
The field $\zeta_\textup{ext}$ is constant because it satisfies the vacuum
Maxwell equation
\begin{equation*}
d\zeta_\textup{ext} = 0
\end{equation*}
inside the charged continuum (consider equation (\ref{maxlor11_d_E}) with
$\rho=0$). The self-field
$\zeta_\textup{self}$ is to be defined in terms of the initial reduced proper
charge density $\rho(0,z)$:
\begin{equation}
\label{maxlor11_zeta-self_and_rho}
\zeta_\textup{self}(z) \equiv \frac{1}{2}
\left[\int^z_{-\infty}\rho(0,s)\,ds - \int^\infty_z\rho(0,s)\,ds\right]
\end{equation}
and satisfies the asymptotic conditions
\begin{align*}
&\underset{z\rightarrow\infty}{\text{lim}}\zeta_\textup{self}(z) = \frac{1}{2}Q,\\
&\underset{z\rightarrow-\infty}{\text{lim}}\zeta_\textup{self}(z) = -\frac{1}{2}Q
\end{align*}
where $Q=\int^\infty_{-\infty}\rho(0,z)dz$ is the total charge per unit area of the continuum.

\begin{figure}
\begin{center}
\scalebox{1.0}{\includegraphics{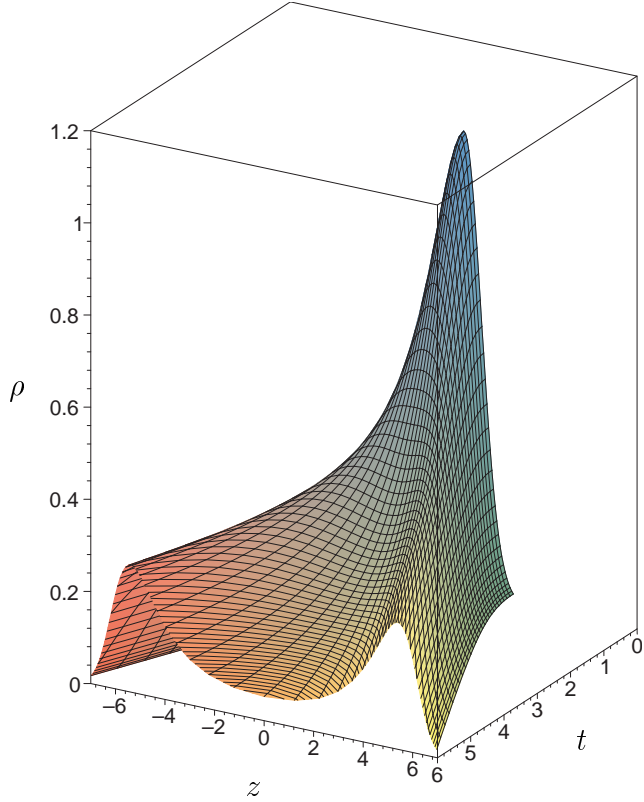}}
\caption{\label{figure:double_wall_3d}The time history of the initial
charge distribution
(\ref{maxlor11_initial_charge_distribution_example}) with zero applied
external field.}
\end{center}
\end{figure}

\begin{figure}
\begin{center}
\scalebox{1.0}{\includegraphics{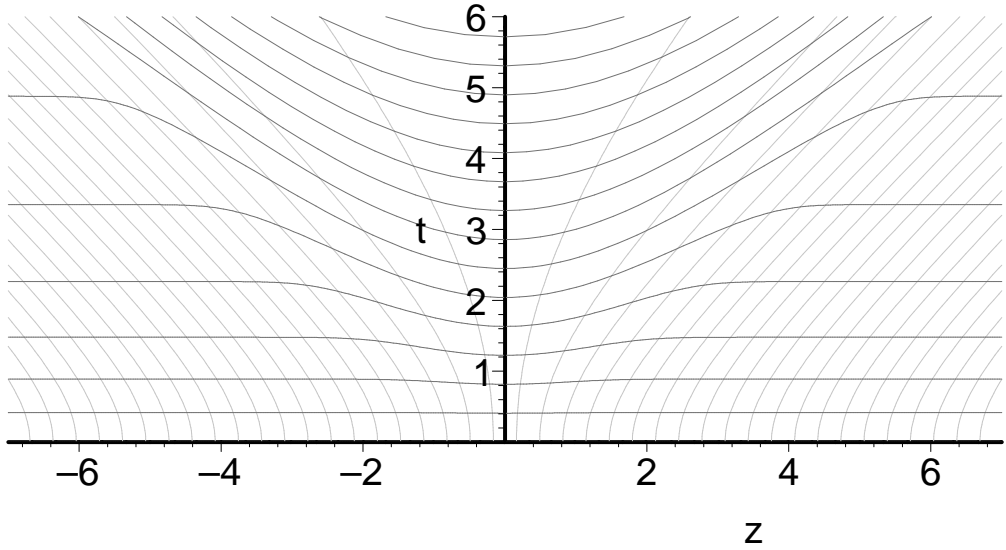}}
\caption{\label{figure:coordinate_lines_double_wall}The lines of constant $\tau$
and $\sigma$ generated by the initial charge distribution
(\ref{maxlor11_initial_charge_distribution_example}) with zero applied
external field. Lines of constant $\tau$, except $\tau=0$, are dark grey and
lines of constant $\sigma$ (the streamlines of the flow), except $\sigma=0$, are light grey.
The black line $\tau=0$ coincides
with the $z$-axis and the black line $\sigma=0$ coincides with the $t$-axis.}
\end{center}
\end{figure}

\begin{figure}
\begin{center}
\scalebox{1.0}{\includegraphics{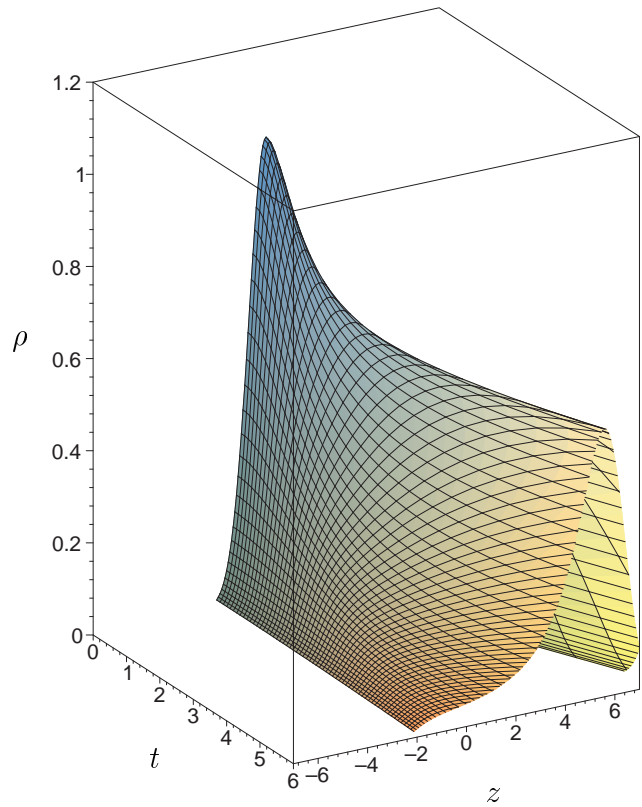}}
\caption{\label{figure:single_wall_3d}The time history of the initial
charge distribution
(\ref{maxlor11_initial_charge_distribution_example}) with finite applied
external field.}
\end{center}
\end{figure}

\begin{figure}
\begin{center}
\scalebox{1.0}{\includegraphics{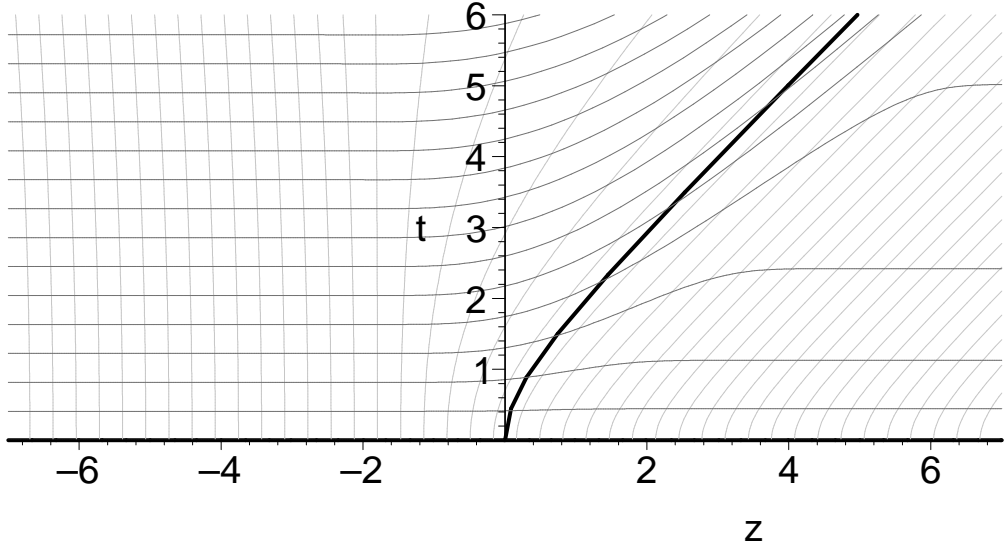}}
\caption{\label{figure:coordinate_lines_single_wall}The lines of constant $\tau$
and $\sigma$ generated by the initial charge distribution
(\ref{maxlor11_initial_charge_distribution_example}) with finite applied
external field. Lines of constant $\tau$, except $\tau=0$, are dark grey and
lines of constant $\sigma$ (the streamlines of the flow), except $\sigma=0$, are light grey.
The black line $\tau=0$ coincides
with the $z$-axis and the black curve lying in the region $z>0$ is $\sigma=0$.}
\end{center}
\end{figure}

Illustrations of the above results are shown in figures
\ref{figure:double_wall_3d}-\ref{figure:coordinate_lines_single_wall}.
In figures \ref{figure:double_wall_3d} and
\ref{figure:coordinate_lines_double_wall} the
external field is set to zero so the charge distribution $\rho$ evolves
entirely under its own self-field. In figures \ref{figure:single_wall_3d} and
\ref{figure:coordinate_lines_single_wall} the external field is set
to a finite value. Figures \ref{figure:coordinate_lines_double_wall}
and \ref{figure:coordinate_lines_single_wall} show the streamlines of
$V$ in spacetime for each case.
The initial charge distribution is taken as the Gaussian
\begin{equation}
\label{maxlor11_initial_charge_distribution_example}
\rho(0,z) = \exp(-z^2)
\end{equation}
and, using (\ref{maxlor11_zeta-self_and_rho}), the associated electric self-field is
\begin{equation*}
\zeta_\textup{self}(z) = \frac{\sqrt{\pi}}{2}\text{erf}(z).
\end{equation*}
The external field $\zeta_\textup{ext}$ is chosen as:
\begin{equation*}
\zeta_\textup{ext} =
\begin{cases}
0\\
0.875.
\end{cases}
\end{equation*}
In the first case (figures \ref{figure:double_wall_3d} and \ref{figure:coordinate_lines_double_wall}) the initial Gaussian ``wall of charge'' evolves into two mirror-image
``walls of charge'' that propagate with equal Newtonian speed in opposite
directions and tend towards the speed of light as $t\rightarrow\infty$.
In the second case (figures \ref{figure:single_wall_3d} and
\ref{figure:coordinate_lines_single_wall}) the initial Gaussian ``wall of charge'' is
accelerated by a combination of the external field and its self-field and
approximately maintains its shape.

\subsection{Laboratory frame description}
Although the solutions above are exact in terms of $(\tau,\sigma)$, solving (\ref{maxlor11_sigma_imp}) for $\hat{\sigma}(t,z)$ in closed
form is impossible if $\zeta$ is arbitrary. One way of tackling this
problem is to introduce a running parameter $\Vep>0$, replace
$\hat{\sigma}$ by a $1$-parameter family of functions $\hat{\sigma}^\Vep$
and solve (\ref{maxlor11_sigma_imp}) for $\hat{\sigma}^\Vep$ order-by-order in
$\Vep$.

To effect the coordinate transformation from $(\tau,\sigma)$ to
$(t,z)$ $\zeta$ is replaced with the $1$-parameter family of functions $\zeta^\Vep$:
\begin{equation}
\label{maxlor11_zeta_Vep_def}
\zeta^\Vep(\sigma) = \frac{1}{\Vep}\zeta_{-1} + \zeta_0(\sigma).
\end{equation}
where $\zeta^\Vep_\textup{ext}=\frac{1}{\Vep}\zeta_{-1}\neq 0$ and
$\zeta_\textup{self}=\zeta_0$. One may interpret
(\ref{maxlor11_zeta_Vep_def}) by saying that
external field effects (the first term $\frac{1}{\Vep}\zeta_{-1}$)
dominate the effects of space-charge (the second term
$\zeta_0$).
The coefficients of the $\Vep$ terms
in (\ref{maxlor11_zeta_Vep_def}) have been labelled by the
$\Vep$-order of the term in which they appear; this convention will be
used consistently throughout this article.

A solution to (\ref{maxlor11_sigma_imp}) can be obtained
order-by-order in $\Vep$ using the series ansatz for $t>0$
\begin{equation}
\label{maxlor11_sigma_Vep_series}
\hat{\sigma}^\Vep(t,z) = \sum\limits^\infty_{n=0}\Vep^n\hat{\sigma}_n(t,z).
\end{equation}

The electric field component $\calE^\Vep(t,z)$ and $4$-velocity field
$V^\Vep$ can be written entirely in terms of $\hat{\sigma}^\Vep$, $t$ and $z$:
\begin{align*}
&\calE^\Vep(t,z) = \zeta^\Vep\left(\hat{\sigma}^\Vep(t,z)\right)\\
&V^\Vep = \sqrt{1+[\zetaVep]^2t^2}\PD_t + \zetaVep t \PD_z.
\end{align*}
Using (\ref{maxlor11_zeta_Vep_def}), (\ref{maxlor11_sigma_imp}),
(\ref{maxlor11_sigma_Vep_series}), the choice\footnote{Similar
solutions are obtained for $\zeta_{-1}<0$ but with $z+t$ replacing
$z-t$.}
$\zeta_{-1}>0$ and $t>0$,
by equating orders in $\Vep$ it follows that
\begin{align}
\label{maxlor11_sigmahat_Vep_series}
&\hat{\sigma}^\Vep(t,z) = z - t +\frac{\Vep}{\zeta_{-1}}
-\frac{1+2t\zeta_0(z-t)}{2t\zeta_{-1}^2}\Vep^2 + O(\Vep^3),\\
\label{maxlor11_E_Vep_series}
&\calE^\Vep(t,z) = \frac{1}{\Vep}\zeta_{-1} + \zeta_0(z-t) +
\frac{\zeta_0^\prime(z-t)}{\zeta_{-1}}\Vep + O(\Vep^2),\\
\label{maxlor11_V_Vep_series}
&V^\Vep = \left[\frac{1}{\Vep}\zeta_{-1} +
\zeta_0(z-t)\right]t\left(\PD_t+\PD_z\right) +
\left[\frac{1+2t^2\zeta_0^\prime(z-t)}{2t\zeta_{-1}}\PD_t +
\frac{\zeta_0^\prime(z-t)t}{\zeta_{-1}}\PD_z\right]\Vep
 + O(\Vep^2)
\end{align}
where $\zeta^\prime_0(z) \equiv \frac{d\zeta_0}{dz}(z)$. Although
$V^\Vep\cdot V^\Vep = -1$ the $(t,z)$ components of $V^\Vep$ diverge in the limits
$t\rightarrow\infty$ and $t\rightarrow 0$. Divergences as
$t\rightarrow\infty$ are expected because the dominant term in
the electric field is constant, and so the charged continuum is
undergoing constant acceleration to leading order in $\Vep$.
Although $V^\Vep$ diverges as $t\rightarrow\infty$, the electric
current $J^\Vep\equiv\rho^\Vep V^\Vep$ is
\begin{equation}
\label{maxlor11_J_Vep_series}
\begin{split}
J^\Vep &= \rho^\Vep V^\Vep = \widedual{\# d{\calE}^\Vep}\\
&= \zeta^\prime_0(z-t)\left(\PD_t+\PD_z\right) +
\Vep\frac{\zeta^{\prime\prime}_0(z-t)}{\zeta_{-1}}\left(\PD_t+\PD_z\right)
+ O(\Vep^2),
\end{split}
\end{equation}
where (\ref{maxlor11_d_E}) has been used. Since $J^\Vep=\rho^\Vep V^\Vep$,
$J^\Vep=O(\Vep^0)$ and $V^\Vep=O(\Vep^{-1})$ it follows that the
reduced proper charge density $\rho^\Vep=O(\Vep)$.
The leading order contribution to $\rho^\Vep$ is obtained by comparing
(\ref{maxlor11_J_Vep_series}) with (\ref{maxlor11_V_Vep_series}) and
using $J_0=\rho_1 V_{-1}$ where
\begin{equation}
J^\Vep = \sum\limits^\infty_{n=0} \Vep^n J_n,\qquad
V^\Vep = \sum\limits^\infty_{n=-1} \Vep^n V_n,\qquad
\rho^\Vep = \sum\limits^\infty_{n=1} \Vep^n \rho_n
\label{maxlor11_J_V_rho_Vep_series}
\end{equation}
It follows that\footnote{It should be noted that the equation
$\rho^\Vep=-J^\Vep\cdot V^\Vep$
cannot be used to calculate (\ref{maxlor11_rho_Vep_series}) without
first calculating $J_2$ since $\rho_1=-(J_2\cdot V_{-1} + J_1\cdot V_0
+ J_0\cdot V_1)$.}
\begin{equation}
\label{maxlor11_rho_Vep_series}
\rho^\Vep = \Vep\frac{\zeta^\prime_0(z-t)}{\zeta_{-1}t} + O(\Vep^2).
\end{equation}

Equation (\ref{maxlor11_J_Vep_series}) shows that $J^\Vep$ is bounded to
$O(\Vep^2)$ as $t\rightarrow\infty$.
The divergence as $t\rightarrow 0$ of the coefficient of $\Vep$ in
equation (\ref{maxlor11_V_Vep_series}) stems from the square-root term in
(\ref{maxlor11_sigma_imp}). Evidently, the domain on which the
approximate solution is valid (the half-plane $t>0$) is a subset of the domain
of the exact solution (the entire $(t,z)$ plane).
This is not too surprising since the charged
continuum is \emph{at rest} at $t=0$ and the leading term $J_0$ in the
series for $J^\Vep$ is light-like:
\begin{equation*}
J_0\cdot J_0 = 0.
\end{equation*}
There is no observer frame in which $J_0$ is instantaneously at rest.

This simple example suggests that the series
\begin{equation*}
F^\Vep = \sum\limits^\infty_{n=-1} \Vep^nF_n,\qquad
V^\Vep = \sum\limits^\infty_{n=-1} \Vep^nV_n,\qquad
\rho^\Vep = \sum\limits^\infty_{n=1} \Vep^n\rho_n
\end{equation*}
where $F_{-1}$ is an external field (a solution to the source-free Maxwell
equations), should be inserted into the field system \Fieldsysrefs
which is then solved order-by-order in $\Vep$.

This approach can lead to the series expansions for the highly symmetric
solutions just discussed, but can also yield
$\Vep$ expansions for solutions that depend on $(x,y)$ as well as
$(t,z)$. A consequence of the above series expansions is that the
electric $4$-current has the form
\begin{equation*}
J^\Vep = \rho^\Vep V^\Vep = \sum\limits^\infty_{n=0} \Vep^n J_n.
\end{equation*}
It will be shown below that the above expansions partially decouple
the field system \Fieldsysrefs
yielding an infinite hierarchy of equations that are amenable to
solution when supplemented with appropriate boundary conditions and initial data.
\section{The perturbation scheme}
\label{section:construction_of_the_scheme}

Based on the above exact symmetric solution it is assumed that
there exists a class of solutions to \Fieldsysrefs that represents
configurations of charged particles in {\em ultra-relativistic}
collective motion.  It is shown in Appendix
\ref{appendix:ultrarel_light-like_defs} that the notion of
an ultra-relativistic 4-velocity vector field can be made precise
and should be distinguished from the magnitude of a relative
Newtonian speed  that can be defined for {\em any two} time-like
4-vector fields. An ultra-relativistic velocity field is a
pointwise limiting concept that depends on the existence of the
forward light-cone structure at each event in spacetime. To
leading order in the expansion defined below the velocity
field of the charged continuum is light-like. The full
series will be considered as an asymptotic expansion for a
solution to \Fieldsysrefs and physically represents an
ultra-relativistic configuration controlled by the parameter
$\varepsilon$. Such configurations cannot of course be
exhaustive. They are chosen to be representative of the class
relevant to charged beams in high-energy accelerators.

Introduce perturbation series for
$(V^\varepsilon,\rho^\varepsilon,F^\varepsilon)$ in $\Vep$ of the
form
\begin{align}
V^\varepsilon=\sum_{n=-1}^\infty
\varepsilon^n V_n
\,,\qquad
\rho^\varepsilon=\sum_{n=1}^\infty
\varepsilon^n \rho_n
\,,\qquad
F^\varepsilon=\sum_{n=-1}^\infty
\varepsilon^n F_n
\label{intro_V_R_F_expan}
\end{align}
where
\begin{align}
V_n\in\Gamma T\man
\,,\qquad
\rho_n\in\Gamma\Lambda^0\man
\,,\qquad
F_n\in\Gamma\Lambda^2\man
\label{intro_V_R_F_memeber}
\end{align}
\begin{theorem}
\label{np_hierarchy_theorem}
Using \Fieldsysintegrefs the coefficients of the expansions in
(\ref{intro_V_R_F_expan}) satisfy
\begin{align}
& d F_{n-1} = 0
\qquad\textup{ for } n\in\Set{0,1,2,\ldots}
\label{np_maxwell_d_F_n}
\\
&
d \star F_{n-1} =
\begin{cases}
0 &\textup{for } n=0
\\
\displaystyle
-
\sum_{r=1}^{n}\star\rho_r \dual{V}_{n-r-1}
&\textup{for } n\in\Set{1,2,\ldots}
\end{cases}
\label{np_maxwell_d_star_F_n}
\\
&
\sum_{r=0}^{n}
\nabla_{V_{r-1}} \dual{V}_{n-r-1}
=
\sum_{r=0}^{n}
i_{V_{r-1}} F_{n-r-1}
\label{np_maxwell_lorentz_V_n}
\qquad\textup{ for } n\in\Set{0,1,2,\ldots}
\\
&
\sum_{r=0}^{n}
{V_{r-1}} \cdot{V}_{n-r-1}
=
\begin{cases}
-1 &\textup{ for } n=2
\\
0 &\textup{ for }n\in\Set{0,1,3,4,5,\ldots}
\end{cases}
\label{np_V_n_dot_V_n}
\\
&
\sum_{r=1}^{n+1}d \star\left(\rho_r \dual{V}_{n-r}\right)
=0
\qquad\textup{ for } n\in\Set{0,1,2,\ldots}
\label{np_d_star_rho_V_n}
\end{align}
The partially decoupled equations
(\ref{np_maxwell_d_F_n}-\ref{np_d_star_rho_V_n}) are amenable to an
ordered analysis and can be arranged into the hierarchy:
\begin{align}
\makebox{\upshape
\begin{tabular}{l}
For $N\in\Set{0,1,2,\ldots}$:
\\
\qquad
Step $3N+1$: Solve (\ref{np_maxwell_d_F_n}) and
(\ref{np_maxwell_d_star_F_n}) with $n=N$ for $F_{n-1}$.
\\
\qquad
Step $3N+2$: Solve (\ref{np_maxwell_lorentz_V_n}) and
(\ref{np_V_n_dot_V_n}) with $n=N$
for $V_{n-1}$.
\\
\qquad
Step $3N+3$: Solve (\ref{np_d_star_rho_V_n}) with $n=N$
for $\rho_{n+1}$.
\end{tabular}
}
\label{np_solution_hierarchy}
\end{align}
\end{theorem}
\begin{proof}
Equations (\ref{np_maxwell_d_F_n}-\ref{np_d_star_rho_V_n}) are obtained by inserting the series (\ref{intro_V_R_F_expan}) into \Fieldsysrefs and the
integrability condition (\ref{np_maxwell_d_rho_star_V}).
The hierarchical structure follows by inspection.
\end{proof}
To illustrate the partially decoupled structure in
(\ref{np_solution_hierarchy}) consider the first 8 steps in
detail:
\begin{steplist}
\item
Adopt an external electromagnetic field $F_{-1}$ i.e. a solution of
the source-free equations
\begin{align}
d F_{-1} = 0
\qquad\text{and}\qquad
d \star F_{-1} = 0.
\label{np_maxwell_F_m1}
\end{align}
\item
\label{np_solution_hierarchy_step_2}
Solve
\begin{align}
\nabla_{V_{-1}} \dual{V}_{-1} = i_{V_{-1}} F_{-1}
\qquad\text{subject to}\qquad
V_{-1}\cdot V_{-1} =0
\label{np_lorentz_V_m1}
\end{align}
for $V_{-1}\neq 0$, where $F_{-1}$ is data
obtained in the previous step.

Although (\ref{np_lorentz_V_m1}) is non-linear in the
unknown field $V_{-1}$ it is straightforward to analyse; it can be
written as a quasi-linear second-order ordinary differential equation
for the integral curves of $V_{-1}$ and is the \emph{only} non-linear
differential equation in this scheme.

It may seem that the pair of equations above
offer $5$ scalar equations for $4$ unknown scalars. However, the
contraction of the first equation with
$V_{-1}$ identically vanishes, so there are only $4$ independent
equations.
This can be seen\footnote{The split of (\ref{np_lorentz_V_m1}) with respect to the
laboratory frame is given in Appendix
\ref{appendix:Gibbs_3-vector_version}. The equation obtained
by contracting (\ref{np_lorentz_V_m1}) with the laboratory observer
field $\PD_t$ is redundant because it follows as a consequence of
(\ref{gibbs_step2_diff}) and (\ref{gibbs_step2_alg}).}
by writing
(\ref{np_lorentz_V_m1}) with respect to an arbitrary basis and noting
that one of the $5$ components of (\ref{np_lorentz_V_m1}) is a
consequence of the other $4$.
\item
The leading order reduced proper charge density $\rho_1$ is a solution to
\noindent
\begin{align}
d \star\left(\rho_1 \dual{V}_{-1}\right) =0.
\label{np_d_star_rho_1}
\end{align}
\item
The $2$-form $F_{0}$ is a solution to Maxwell's equations with the current
$\rho_1\dual{V}_{-1}$ as a source:
\begin{align}
d F_{0} = 0
\qquad\text{and}\qquad
d \star F_{0} = -\star \rho_1\dual{V}_{-1}.
\label{np_maxwell_F_0}
\end{align}
The conservation equation (\ref{np_d_star_rho_1}) ensures that
(\ref{np_maxwell_F_0}) has a solution.
\item
The zero order velocity field $V_{0}$ is obtained from the
\emph{linear} equations
\begin{align}
\nabla_{V_{-1}} \dual{V}_{0} +
\nabla_{V_{0}} \dual{V}_{-1} =
i_{V_{-1}} F_{0} +
i_{V_{0}} F_{-1}
\qquad\text{subject to}\qquad
V_{-1}\cdot V_{0} =0.
\label{np_lorentz_V_0}
\end{align}
\item
The second order coefficient $\rho_2$ in the $\Vep$ expansion for
the reduced proper charge density $\rho^\Vep$ satisfies
\noindent
\begin{align}
d\star\left(\rho_2 \dual{V}_{-1}\right) + d\star\left(\rho_1 \dual{V}_{0}\right) =0.
\label{np_d_star_rho_2}
\end{align}
\item
Equation (\ref{np_d_star_rho_2}) ensures that
\begin{align}
d F_{1} = 0
\qquad\text{and}\qquad
d \star F_{1} =
- \star\rho_2 \dual{V}_{-1} - \star\rho_1 \dual{V}_{0}
\label{np_maxwell_F_1}
\end{align}
may be solved for the coefficient $F_1$ (the first order term in
the $\Vep$ expansion of $F^\Vep$) given appropriate initial and
boundary conditions. \item The linear partial differential
equation for $V_{1}$ is
\begin{align}
\hspace{-2em}
\nabla_{V_{-1}} \dual{V}_{1} +
\nabla_{V_{0}} \dual{V}_{0} +
\nabla_{V_{1}} \dual{V}_{-1} =
i_{V_{-1}} F_{1} +
i_{V_{0}} F_{0} +
i_{V_{1}} F_{-1}
\qquad\text{subject to}\qquad
2V_{-1}\cdot V_{1} + V_{0}\cdot V_{0} =-1.
\label{np_lorentz_V_1}
\end{align}
\end{steplist}

$V^\Vep$ must be
calculated to at least first order in $\Vep$ to obtain a time-like
vector field. To see this let $X^\Vep=\Vep^{-1}V_{-1} + V_0$ and note $X^\Vep\cdot X^\Vep
= V_0\cdot V_0$ using the metric-product conditions in
(\ref{np_lorentz_V_m1}) and (\ref{np_lorentz_V_0}). Using lemma
\ref{lm_Vm1_ne_0} in Appendix \ref{appendix:ultrarel_light-like_defs} $V_0\cdot V_0\ge 0$ and so  $X^\Vep$ is light-like or
space-like. Now introduce $Y^\Vep = X^\Vep+\Vep V_1 = \Vep^{-1}V_{-1} + V_0 +
\Vep V_1$ and note $Y^\Vep\cdot Y^\Vep = V_0\cdot V_0 + 2V_1\cdot
V_{-1} + O(\Vep) = -1 + O(\Vep)$ using
(\ref{np_lorentz_V_1}). Therefore, there exists some value of $\Vep>0$
for which $Y^\Vep$ is time-like.

It has already been stated that the pair of equations
(\ref{np_lorentz_V_m1}) in step \ref{np_solution_hierarchy_step_2} may
be considered as $4$ non-linear equations in $4$ unknowns. Furthermore:
\begin{lemma}
\label{lm_Vm1_proj_0} The equations in each step $3N+2$, for $N\ge1$,
are equivalent to a set of $4$ independent inhomogeneous linear
equations for the $4$ unknown components of $V_{N-1}$.
\end{lemma}

\begin{proof}
Step $3N+2$ involves solving (\ref{np_maxwell_lorentz_V_n}) and
(\ref{np_V_n_dot_V_n}) for $V_{N-1}$.
Assume that the equations in steps $1$ to $3N+1$ have
been solved and define $L_n$ as
\begin{align}
\label{lm_def_L_n}
L_n = \sum_{r=0}^{n}
\left(
\nabla_{V_{n-r-1}} \dual{V}_{r-1}
-
i_{V_{r-1}} F_{n-r-1}
\right).
\end{align}
Clearly (\ref{np_maxwell_lorentz_V_n}) is equivalent to
$L_n=0$. The vector fields $V_{-1},\ldots,V_{N-2}$ have been obtained in steps $1$
to $3N+1$ and so $L_n=0$ for $n=0,\ldots,N-1$. It will now be shown
that given any $V_{N-1}$ satisfying (\ref{np_V_n_dot_V_n}) with $n=N$
the scalar field $i_{V_{-1}}L_N$ identically vanishes.

From (\ref{np_V_n_dot_V_n}) for $n=0,\ldots,N$
\begin{align*}
0 =
\nabla_{V_s}\left(\sum_{r=0}^{n}  {V_{r-1}} \cdot{V}_{n-r-1}\right)
= &
\sum_{r=0}^{n} (\nabla_{V_s}V_{r-1})\cdot{V}_{n-r-1}
+
\sum_{r=0}^{n} V_{r-1}\cdot \nabla_{V_s}{V}_{n-r-1}\\
= &
\sum_{r=0}^{n} (\nabla_{V_s}V_{r-1})\cdot{V}_{n-r-1}
+ \sum_{q=n}^{0} V_{n-q-1}\cdot \nabla_{V_s}{V}_{q-1}
\end{align*}
where the second sum has been
re-labelled by $q=n-r$. The two sums are identical so
\begin{align}
\sum_{r=0}^{n} (\nabla_{V_s}V_{n-r-1})\cdot{V}_{r-1} = 0.
\label{lm_sums_nab_VV_0}
\end{align}
Using (\ref{lm_def_L_n}) and $L_n=0$ for all $0\le n<N$
\begin{align}
i_{V_{-1}} L_{N}
= &
\sum_{n=0}^N i_{V_{n-1}} L_{N-n}
\notag
\\
= &
\sum_{n=0}^N i_{V_{n-1}}
\sum_{r=0}^{N-n}
\left(
\nabla_{V_{N-n-r-1}} \dual{V}_{r-1}
-
i_{V_{r-1}} F_{N-n-r-1}
\right)
\notag
\\
=&
\sum_{n=0}^N
\sum_{r=0}^{N-n}
\left(
{V_{n-1}}
\cdot
\nabla_{V_{N-n-r-1}}
{V_{r-1}}
-
i_{V_{n-1}}
i_{V_{r-1}} F_{N-n-r-1}
\right)
\notag
\\
=&
\sum_{n=0}^N
\sum_{s=n}^N
\left(
{V_{n-1}}
\cdot
\nabla_{V_{N-s-1}}
{V_{s-n-1}}
-
i_{V_{n-1}}
i_{V_{s-n-1}} F_{N-s-1}
\right)
\label{lm_sums_nab_VV_relabel}
\end{align}
where, in the final step, the innermost sum has been re-labelled by
$s=n+r$. The double summation in (\ref{lm_sums_nab_VV_relabel}) is over the set
\begin{equation*}
\Set{\strut(n,s)\, |\, n,s\in\Set{0,1,\ldots,N}, n\le s}
\end{equation*}
and so
\begin{equation*}
\sum_{n=0}^N\sum_{s=n}^N S_{ns} = \sum_{s=0}^N\sum_{n=0}^s S_{ns}
\end{equation*}
for any summand $S_{ns}$. Therefore, using
(\ref{lm_sums_nab_VV_relabel})
\begin{align}
i_{V_{-1}} L_{N}
& =
\sum_{s=0}^N
\left(
\sum_{n=0}^s
{V_{n-1}}
\cdot
\nabla_{V_{N-s-1}}
{V_{s-n-1}}
-
\sum_{n=0}^s
i_{V_{n-1}}
i_{V_{s-n-1}} F_{N-s-1}
\right)
\notag
\\
&= 0
\label{lm_i_Vm1_L_N_0}
\end{align}
where the first term in the summand vanishes because of
(\ref{lm_sums_nab_VV_0}) and the second term vanishes because $i_{V_{n-1}}
i_{V_{s-n-1}} = - i_{V_{s-n-1}}i_{V_{n-1}}$ and the sum over $s$ is
from $0$ to $n$.

Note that the lemma applies to step $5$ and beyond and so, for the
purposes of the lemma, the fields
$F_{-1},V_{-1},\rho_1,F_0$ obtained in steps $1$ to $4$ are prescribed
data.
Let $\Set{X_1,X_2,X_3,X_4}$, with $X_i\in\Gamma T\man$ for
$i=1,2,3,4$, be any vector
frame such that $X_1=V_{-1}$
and define $\Set{e^1,e^2,e^3,e^4}$, where
$e^i\in\Gamma\Lambda^1\man$, to be the coframe dual to
$\Set{X_1,X_2,X_3,X_4}$:
\begin{equation*}
e^i(X_j) = \delta^i_j
\end{equation*}
where $\delta^i_j$ is the Kronecker delta. Equation
(\ref{np_maxwell_lorentz_V_n}) can be written as
\begin{align}
\calL(\dualV_{N-1})=\xi_{N-1}
\label{np_L_V_n_xi}
\end{align}
where $\calL$ is the $\Real$-linear differential operator on
$\Gamma\Lambda^1\man$
\begin{align*}
\calL(\dualV_{N-1}) =
\nabla_{V_{-1}} \dualV_{N-1} +
\nabla_{V_{N-1}} \dualV_{-1} +
i_{V_{N-1}} F_{-1}
\end{align*}
where $\xi_{N-1}\in\Gamma\Lambda^1\man$ depends on
$\Set{V_{-1},V_0\ldots,V_{N-2},F_{0},\ldots,F_{N-1}}$ but is
independent of $V_{N-1}$. Hence, using (\ref{lm_i_Vm1_L_N_0})
\begin{align}
i_{X_1}(\calL(\dualV_{N-1})-\xi_{N-1})=
i_{V_{-1}}(\calL(\dualV_{N-1})-\xi_{N-1})=0
\label{np_X_1_L_V_n_xi}
\end{align}
Likewise, equation (\ref{np_V_n_dot_V_n}) can be expressed as
\begin{align}
i_{X_1}\dualV_{N-1} = i_{V_{-1}}\dualV_{N-1} = C_{N-1}
\label{np_X_dot_V_n_1}
\end{align}
where the scalar field $C_{N-1}\in\Gamma\Lambda^0\man$ depends only on
$\Set{V_0,\ldots,V_{N-2}}$.

Using the coframe $\Set{e^1,e^2,e^3,e^4}$
\begin{align*}
\dualV_{N-1}=\sum_{i=1}^4 \lambda_{N-1,i} e^i
\end{align*}
for component fields $\lambda_{N-1,i}$. From
(\ref{np_X_dot_V_n_1}) $\lambda_{N-1,1}=C_{N-1}$ and
(\ref{np_L_V_n_xi}) becomes
\begin{align}
\calL\Big(\sum_{i=2}^4 \lambda_{N-1,i} e^i\Big)
=
\xi_{N-1} -\calL(C_{N-1} e^1).
\label{np_L_lambda_e_n_xi}
\end{align}
From (\ref{np_X_1_L_V_n_xi}) $i_{X_1}$ annihilates
(\ref{np_L_lambda_e_n_xi}) identically and
\begin{align*}
i_{X_j}\left(\calL\Big(\sum_{i=2}^4 \lambda_{N-1,i} e^i\Big)\right)
=
i_{X_j}\left(\xi_{N-1} -\calL(C_{N-1} e^1)\right)
\,\qquad\text{for}\qquad
j=2,3,4.
\end{align*}
Thus, using the inhomogeneous linear algebraic constraint
(\ref{np_V_n_dot_V_n}), $3$ inhomogeneous linear differential
equations have been obtained for the $3$ unknown fields
$\Set{\lambda_{N-1,2},\lambda_{N-1,3},\lambda_{N-1,4}}$ and
$\lambda_{{N-1},1}$ is obtained using (\ref{np_V_n_dot_V_n}), determining
$V_{N-1}$.
\end{proof}
In conclusion, the only non-linear equations
in the hierarchy developed in
theorem \ref{np_hierarchy_theorem} are those for $V_{-1}$.
\emph{All} of the remaining equations are linear and therefore, in
principle, simpler to solve.

\section{Example : Charged beam propagating in free space}
As a simple application of the hierarchy
(\ref{np_maxwell_d_F_n}-\ref{np_d_star_rho_V_n}) consider
a high-energy charged beam propagating in free space. Unlike the
exact ``wall of charge'' solutions derived earlier, the charge
distributions considered here have finite extent transverse to and
along the direction of propagation.
The first eight steps of the hierarchy
(\ref{np_maxwell_d_F_n}-\ref{np_d_star_rho_V_n}) are:
\begin{steplist}
\item The leading order Maxwell field $F_{-1}$ (the external field)
is chosen to vanish so (\ref{np_maxwell_F_m1}) is trivially satisfied.
\item Using (\ref{np_lorentz_V_m1}) the leading order $4$-velocity field
$V_{-1}$ satisfies
\begin{equation}
\label{egbeam_light-like_geodesic_V_field}
\nabla_{V_{-1}}V_{-1}=0,\quad V_{-1}\cdot V_{-1} = 0
\end{equation}
since the external field $F_{-1}=0$. Therefore, consider
solutions to (\ref{np_maxwell_d_F_n}-\ref{np_d_star_rho_V_n})
adapted to a null geodesic coordinate system $(u,v,x,y)$
where $u=z-t$, $v=z+t$ and
\begin{align}
\notag
g &= - dt\otimes dt + dx\otimes dx + dy\otimes dy + dz\otimes dz\\
\label{egbeam_metric}
&= \frac{1}{2}\left(du\otimes dv + dv\otimes du\right) + dx\otimes
dx + dy\otimes dy
\end{align}
with the spacetime volume $4$-form
\begin{equation*}
\star 1 = dt\wedge dz \wedge \#_\perp 1
\end{equation*}
where
\begin{equation*}
\#_\perp 1 \equiv dx\wedge dy.
\end{equation*}
The pair $(\PD_t,\PD_z)$ and their duals $(dt,dz)$ are related to
$(\PD_v,\PD_u)$ and $(dv,du)$ as follows:
\begin{align*}
&\PD_v = \frac{1}{2}\left(\PD_z+\PD_t\right),\quad\PD_u = \frac{1}{2}\left(\PD_z-\PD_t\right),\\
&dv = dz + dt,\quad du = dz - dt.
\end{align*}
Metric identities used in the subsequent steps are
\begin{equation}
\label{egbeam_metric_identities}
\star(du\wedge\alpha) = (-1)^p du\wedge\#_\perp \alpha,\quad
\star(dv\wedge\alpha) =
-(-1)^p dv\wedge\#_\perp \alpha
\end{equation}
where $\#_\perp(\alpha\wedge\tilde X) = i_X\#_\perp\alpha$ and $\alpha$
is any $p$-form such that $i_{\PD_u}\alpha =
i_{\PD_v}\alpha = 0$.
The vector field $V_{-1}$ is chosen as
\begin{align}
\notag
V_{-1} &= \gamma_{-1}\left(\PD_t+\PD_z\right)\\
\label{egbeam_V_field_-1}
&= 2\gamma_{-1}\PD_v
\end{align}
where $\gamma_{-1}\neq 0$ is assumed and, using
$\nabla\PD_v=\nabla\PD_u=0$, the only non-zero component of
(\ref{egbeam_light-like_geodesic_V_field}) is
\begin{equation}
\label{egbeam_PDv_gamma_-1}
\PD_v\gamma_{-1} = 0.
\end{equation}
Hence, $\gamma_{-1}$  is independent of $v$.
\item
Using (\ref{egbeam_V_field_-1}), equation (\ref{np_d_star_rho_1}) is
\begin{equation*}
\PD_v(\rho_1\gamma_{-1})=0
\end{equation*}
and so, using (\ref{egbeam_PDv_gamma_-1})
\begin{equation*}
\PD_v\rho_1=0.
\end{equation*}
It follows that $\rho_1$ is independent of $v$.
\item\label{egbeam_hierarchy_step_F_0}
Using (\ref{egbeam_metric_identities}), \emph{particular} solutions to
(\ref{np_maxwell_F_0})
may be written
\begin{equation}
\label{egbeam_faraday_0}
F_0 = d\Phi_0\wedge du
\end{equation}
where the scalar potential $\Phi_0$ is chosen to satisfy
\begin{align}
\label{egbeam_potential_equations_0}
\PD_v\Phi_0=0,\quad d_\perp\#_\perp d_\perp\Phi_0 = \rho_1\gamma_{-1}\#_\perp 1
\end{align}
and $d_\perp$ is the exterior derivative in the $(x,y)$ plane defined by
\begin{equation*}
d_\perp f = \PD_x f\, dx + \PD_y f\, dy
\end{equation*}
for any $0$-form $f\in\Gamma \Lambda^0\man$.
\item Using (\ref{np_lorentz_V_0}), the equations for $V_0$ are
\begin{align*}
\nabla_{V_{-1}}\tilde V_0 + \nabla_{V_0}\tilde V_{-1} = i_{V_{-1}}F_0,
\quad V_{-1}\cdot V_0 = 0
\end{align*}
and are solved by the \emph{particular} solution
\begin{equation*}
V_0=0
\end{equation*}
since $i_{V_{-1}}F_0=0$ using
(\ref{egbeam_faraday_0}), (\ref{egbeam_potential_equations_0}) and
(\ref{egbeam_V_field_-1}).
\item
Since $V_0=0$ it follows that (\ref{np_d_star_rho_2}) has the form
\begin{equation*}
d\star\left(\rho_2\tilde V_{-1}\right) = 0
\end{equation*}
and, using (\ref{egbeam_V_field_-1}) and (\ref{egbeam_PDv_gamma_-1}),
is solved by any $\rho_2$ satisfying
\begin{equation*}
\PD_v\rho_2 = 0
\end{equation*}
i.e. $\rho_2$ is independent of $v$.
\item
Since $V_0=0$ the Maxwell equations
(\ref{np_maxwell_F_1}) are
\begin{align*}
dF_1 = 0,\quad d\star F_1 = -\rho_2\star\tilde V_{-1}
\end{align*}
and, employing the same method used in step
\ref{egbeam_hierarchy_step_F_0}, are solved by the \emph{particular} solution
\begin{equation}
\label{egbeam_faraday_1}
F_1 = d\Phi_1\wedge du
\end{equation}
where $\Phi_1$ is chosen to satisfy
\begin{align}
\label{egbeam_potential_equations_1}
\PD_v\Phi_1=0,\quad d_\perp\#_\perp d_\perp\Phi_1 = \rho_2\gamma_{-1}\#_\perp 1.
\end{align}
\item
Using (\ref{np_lorentz_V_1}), the equations for $V_1$ are
\begin{equation}
\label{egbeam_lorentz_V_1}
\nabla_{V_{-1}}\tilde V_1 + \nabla_{V_1}\tilde V_{-1} =
i_{V_{-1}}F_1,\quad V_{-1}\cdot V_1 = -\frac{1}{2}
\end{equation}
and since $i_{V_{-1}}F_1=0$ (which follows using (\ref{egbeam_faraday_1}),
(\ref{egbeam_potential_equations_1}) and (\ref{egbeam_V_field_-1})) a \emph{particular} solution to
(\ref{egbeam_lorentz_V_1}) is
\begin{equation*}
V_1 = -\frac{1}{2\gamma_{-1}}\PD_u,\quad \PD_u \gamma_{-1} = 0
\end{equation*}
using (\ref{egbeam_PDv_gamma_-1}), (\ref{egbeam_V_field_-1}) and
$\nabla\PD_v=\nabla\PD_u=0$.
Hence, the function $\gamma_{-1}$ is independent of $u$ and therefore,
using (\ref{egbeam_PDv_gamma_-1}), $\gamma_{-1}$ depends only on $x$ and $y$.
\end{steplist}
In summary these solutions, describing an ultra-relativistic charged distribution
propagating along the $z$-axis with its electromagnetic self-fields, are
\begin{align*}
\begin{split}
&V^\Vep = \frac{1}{\Vep}2\gamma_{-1}\PD_v -
\Vep\frac{1}{2\gamma_{-1}}\PD_u + O(\Vep^2)\\
&\quad\, =
\left(\frac{1}{\Vep}\gamma_{-1}+\frac{\Vep}{4\gamma_{-1}}\right)\PD_t
+
\left(\frac{1}{\Vep}\gamma_{-1}-\frac{\Vep}{4\gamma_{-1}}\right)\PD_z
+ O(\Vep^2)
\end{split}\\
\begin{split}
&F^\Vep = d\Phi_0\wedge du + \Vep d\Phi_1\wedge du + O(\Vep^2)\\
&\quad\, = -(d\Phi_0 + \Vep d\Phi_1)\wedge dt + (d\Phi_0 + \Vep
d\Phi_1)\wedge dz + O(\Vep^2)\\
\end{split}
\end{align*}
where $\Phi_0$ and $\Phi_1$ satisfy
\begin{align}
\label{egbeam_Poisson_eqns}
&d_\perp\#_\perp d_\perp\Phi_0 = \gamma_{-1}\rho_1\#_\perp 1,\quad
d_\perp\#_\perp d_\perp\Phi_1 = \gamma_{-1}\rho_2\#_\perp 1
\end{align}
and
\begin{align*}
&\Phi_0 = \hat{\Phi}_0(u,x,y),\quad \Phi_1 = \hat{\Phi}_1(u,x,y),\\
&\rho_1 = \hat{\rho}_1(u,x,y),\quad \rho_2 = \hat{\rho}_2(u,x,y),\\
&\gamma_{-1} = \hat{\gamma}_{-1}(x,y)
\end{align*}
with $u=z-t$. The scalar fields $\rho_0, \rho_1, \gamma_{-1} $ are
determined by their values on the
space-like hypersurface $t=0$ given as data. The potentials $\Phi_0$
and $\Phi_1$ are then solved in terms of $\gamma_{-1}\rho_0$ and
$\gamma_{-1}\rho_1$ using the $2$-dimensional Poisson
equations (\ref{egbeam_Poisson_eqns}) in the $(x,y)$ plane.

The $3$-velocity of the beam is along the direction $z$ in the
laboratory frame with Newtonian speed
\begin{equation*}
\frac{\frac{1}{\Vep}\gamma_{-1}-\Vep\frac{1}{4\gamma_{-1}}+O(\Vep^2)}{\frac{1}{\Vep}\gamma_{-1}+\Vep\frac{1}{4\gamma_{-1}}+O(\Vep^2)}
= 1 - \frac{\Vep^2}{2\gamma_{-1}^2}+O(\Vep^3).
\end{equation*}
For example, consider a Gaussian
bunch with transverse radius $R_0$ travelling at constant Newtonian
speed $1-\frac{\Vep^2}{2b_0^2}$ to order $\Vep^2$:
\begin{align*}
\hat{\gamma}_{-1}(x,y) = b_0,\quad\hat{\rho}_1(z,x,y) =
a_0\exp\left(-\frac{x^2+y^2}{R_0^2}\right)\Xi(z)
\end{align*}
where $a_0, R_0$ and $b_0$  are constants and $\Xi:\Real\rightarrow\Real$
is a smooth bump function vanishing outside the interval $(-z_1,z_1)$
and $\Xi(z)=1$ for $z\in(-z_2,z_2)$ and $z_1 > z_2 > 0$.
Then the laboratory reduced charge density
$\gamma_{-1}\rho_1$ for some range of $t$ is
\begin{equation*}
\gamma_{-1}\rho_1 = \hat{\gamma}_{-1}(x,y)\hat{\rho}_1(z-t,x,y) = a_0\,b_0\exp\left(-\frac{x^2+y^2}{R^2}\right)\Xi(z-t).
\end{equation*}
Working in the cylindrical polar coordinates $(t,R,\phi,z)$ where
$x=R\cos\phi$ and $y=R\sin\phi$, a
cylindrically symmetric solution to (\ref{egbeam_Poisson_eqns})
well-behaved at $R=0$ is
\begin{equation*}
\Phi_0 = \left\{\int^R_0 a_0\,b_0\, \frac{R_0^2}{2 s}\left[1-\exp\left(-\frac{s^2}{R_0^2}\right)\right]\,ds\right\}\,\Xi(z-t)
\end{equation*}
and the corresponding electromagnetic $2$-form $F_0$ is
\begin{equation*}
F_0 = a_0\,b_0\, \frac{R_0^2}{2
R}\left[1-\exp\left(-\frac{R^2}{R_0^2}\right)\right]\Xi(z-t)\, dR\wedge (-dt + dz).
\end{equation*}
The laboratory electric field is radial, the magnetic field is
azimuthal and their magnitudes are equal and vanish outside
of the support of $\Xi$.
\section{Conclusion}
When analysing the dynamics of charged particle beams it is often
fruitful to adopt a description based on classical fields rather than
classical point particles. Pathologies (such as pre-acceleration)
associated with radiating point particles are avoided by relying on
field-theoretical notions.

A novel analysis of charged beam dynamics has been presented and a
model of a freely propagating charged bunch discussed. The
approach relies on an asymptotic series representation of
solutions to self-consistent spacetime covariant field equations
for $(V,\rho,F)$ describing a charged continuum. The asymptotic
series for $V$  is based on the light-like vector field $V_{-1}$
leading to an ultra-relativistic approximation. The hierarchy of
equations obtained are more amenable to analysis than the original
non-linear field system and particular solutions have been
presented.

There are numerous avenues for the development of this work
involving  ultra-relativistic charged beams in the vicinity of
beam pipes, RF cavities, spoilers, etc. leading to dynamical
effects that are often described in terms of
``wake-fields''~\cite{bane_etal:1984, heifets_kheifets:1991}. This
work will lead to a clearer understanding of radiation-reaction
exhibited by continuum models of charged particle beams.

\section*{Acknowledgements}
The authors  acknowledge  support from the Cockcroft Institute and
the EU GIFT project (NEST- Adventure Project no. 5006) and are
grateful to Professor M Poole for enlightening discussions.

\newcommand{\Newtvel}{{\mathbb V}}
\begin{appendix}
\section{Appendix}
\subsection{Definition of ultra-relativistic and light-like limited
vectors}
\label{appendix:ultrarel_light-like_defs}
The spacetime description of a classical point particle with mass
$m_0>0$ invokes a parametrised curve (world-line) with a
future-pointing time-like tangent vector. For affinely
parametrised curves the tangent vector $V$ (4-velocity) is
normalised: $V\cdot V=-1$. A spacetime reference frame may be
associated with a time-like vector field $U$ with $U\cdot U=-1$ and an
``observer'' in such a frame is modelled by an integral
curve of $U$. The  4-momentum of the particle is defined to be
$p=m_0\,V$. At an event where an observer curve intersects the
world-line of the particle one has the orthogonal decomposition:
\begin{align*}
p={\cal E} U + {\bf P}
\end{align*}
where $U\cdot {\bf P}=0$. Relative to $U$, the scalar ${\cal E}$
is the energy of the particle (in units with $c=1$) and  ${\bf P}$
is its 3-momentum. The Newtonian velocity ${\bf v}$ of the particle relative
to $U$ is
\begin{align}
{\bf v}
=\frac{{\bf P}}{{\cal E}}= -\frac{V}{V\cdot U}-U
\label{appx_def_newt_vel}
\end{align}
Since $U\cdot U = V\cdot V = -1$ implies ${\cal E}^2={\bf P} \cdot {\bf P} +
m_0^2$ it is traditional to say that the particle is
``relativistic'' relative to $U$ if ${\cal E}^2\simeq {\bf P} \cdot
{\bf P}$ or equivalently ${\bf v} \cdot {\bf v}\simeq 1$.
Clearly, such a notion depends on both $V$ and $U$ and the nearness
of $\frac{{\bf P} \cdot {\bf P}}{{\cal E}^2}$ to unity.
Furthermore, one may always find a frame in which ${\bf P}=0$ and
${\cal E}=m_0$. However, given a {\em particular} frame $U$ one may
contemplate a 4-velocity $V$ that is relativistic relative to it.
To generalise this notion for {\em any} frame one may consider a
family of 4-velocities $V^\varepsilon$ and invoke properties of
$V^\varepsilon$ as $\varepsilon$ tends to some limit.

Let
$\varepsilon\in(0,\epsmax)=\Set{\varepsilon\in\Real|0<\varepsilon<\epsmax}$
be the running parameter introduced in the main body of the article.
A $1$-parameter time-like vector field (which is not necessarily normalised)
$W^\varepsilon$ is
\defn{ultra-relativistic} if for \emph{any} time-like vector field $Z\in\Gamma
T\man$
\begin{align}
\lim_{\varepsilon\to0}
\left(
\frac{
\Snorm{W^\varepsilon \cdot Z}
}{
\Vnorm{W^\varepsilon}
}
\right)
=
\infty
\label{intro_def_ultra_rel}
\end{align}
where $\Vnorm{X}\equiv\sqrt{-X\cdot X}$ for any time-like (or
light-like) vector field $X\in\Gamma T\man$.
Note that in this definition $Z$ is \emph{not} a $1$-parameter vector field.
\begin{lemma}
Our definition of an ultra-relativistic vector field
(\ref{intro_def_ultra_rel}) on spacetime is independent of the choice
of the time-like vector field $Z$.
\end{lemma}
\begin{proof}
Clearly (\ref{intro_def_ultra_rel}) is independent of any rescaling of
$Z$. Therefore let $Z$ and $\hZ$ be two $g$-normalised future pointing
time-like vector fields (i.e. $Z\cdot Z=\hat{Z}\cdot\hat{Z}=-1)$
and define the vector field $X_1$ where
\begin{align*}
&X_1=\frac{(\hZ+(\hZ\cdot Z) Z)}{((\hZ\cdot Z)^2-1)^{1/2}}.
\end{align*}
Hence
\begin{align}
&\hZ = -(\hZ\cdot Z) Z + ((\hZ\cdot Z)^2-1)^{1/2} X_1
\label{intro_lm_hZ_Z_X1}
\end{align}
and $X_1\cdot Z=0$, $X_1\cdot X_1=1$. Choose space-like unit vector
fields $X_2,X_3$ so that $(Z,X_1,X_2,X_3)$ is a $g$-orthonormal
frame. It follows that
\begin{align*}
W^\varepsilon\cdot W^\varepsilon
=-(W^\varepsilon\cdot Z)^2 +(W^\varepsilon\cdot X_1)^2
+(W^\varepsilon\cdot X_2)^2 +(W^\varepsilon\cdot X_3)^2
\end{align*}
which implies
\begin{align*}
\frac{(W^\varepsilon\cdot X_1)^2}{\Vnorm{W^\varepsilon}^2}
=
\frac{(W^\varepsilon\cdot Z)^2}{\Vnorm{W^\varepsilon}^2} - 1
-\frac{(W^\varepsilon\cdot X_2)^2}{\Vnorm{W^\varepsilon}^2}
-\frac{(W^\varepsilon\cdot X_3)^2}{\Vnorm{W^\varepsilon}^2}
\end{align*}
and hence
\begin{align}
\frac{\Snorm{W^\varepsilon\cdot X_1}}{\Vnorm{W^\varepsilon}}
\le
\frac{\Snorm{W^\varepsilon\cdot Z}}{\Vnorm{W^\varepsilon}}.
\label{intro_VX1_le_VZ}
\end{align}
The metric contraction of (\ref{intro_lm_hZ_Z_X1}) with
$\frac{W^\varepsilon}{\Vnorm{W^\varepsilon}}$ is
\begin{align*}
\frac{\hZ\cdot W^\varepsilon}{\Vnorm{W^\varepsilon}}
=
-(\hZ\cdot Z)
\frac{Z\cdot W^\varepsilon}{\Vnorm{W^\varepsilon}}
+
((\hZ\cdot Z)^2-1)^{1/2}
\frac{X_1\cdot W^\varepsilon}{\Vnorm{W^\varepsilon}}
\end{align*}
and so
\begin{align*}
(\hZ\cdot Z)
\frac{Z\cdot W^\varepsilon}{\Vnorm{W^\varepsilon}}
= -\frac{\hZ\cdot W^\varepsilon}{\Vnorm{W^\varepsilon}}
+
((\hZ\cdot Z)^2-1)^{1/2}
\frac{X_1\cdot W^\varepsilon}{\Vnorm{W^\varepsilon}}
\end{align*}
which, using the triangle inequality for scalars, implies
\begin{align*}
\Snorm{\hZ\cdot Z}
\frac{\Snorm{Z\cdot W^\varepsilon}}{\Vnorm{W^\varepsilon}}
\le
\frac{\Snorm{\hZ\cdot W^\varepsilon}}{\Vnorm{W^\varepsilon}}
+
((\hZ\cdot Z)^2-1)^{1/2}
\frac{\Snorm{X_1\cdot W^\varepsilon}}{\Vnorm{W^\varepsilon}}.
\end{align*}
Rearranging the above equation gives
\begin{align*}
\frac{\Snorm{\hZ\cdot W^\varepsilon}}{\Vnorm{W^\varepsilon}}
\ge
\Snorm{\hZ\cdot Z}
\frac{\Snorm{Z\cdot W^\varepsilon}}{\Vnorm{W^\varepsilon}}
-
((\hZ\cdot Z)^2-1)^{1/2}
\frac{\Snorm{X_1\cdot W^\varepsilon}}{\Vnorm{W^\varepsilon}}
\end{align*}
and so, using (\ref{intro_VX1_le_VZ})
\begin{align*}
\frac{\Snorm{\hZ\cdot W^\varepsilon}}{\Vnorm{W^\varepsilon}}
\ge
\Snorm{\hZ\cdot Z}
\frac{\Snorm{Z\cdot W^\varepsilon}}{\Vnorm{W^\varepsilon}}
-
((\hZ\cdot Z)^2-1)^{1/2}
\frac{\Snorm{Z\cdot W^\varepsilon}}{\Vnorm{W^\varepsilon}}
=
\frac{\Snorm{Z\cdot W^\varepsilon}}{\Vnorm{W^\varepsilon}}
\left(
\Snorm{\hZ\cdot Z} -
((\hZ\cdot Z)^2-1)^{1/2}
\right).
\end{align*}
Since
$\Snorm{\hZ\cdot Z} -
((\hZ\cdot Z)^2-1)^{1/2} > 0$, and $Z$ and $\hat{Z}$ are independent of $\varepsilon$, it
follows that
\begin{equation*}
\lim_{\Vep\to 0}
\frac{\Snorm{Z\cdot W^\varepsilon}}{\Vnorm{W^\varepsilon}}\to\infty\implies
\lim_{\Vep\to 0}
\frac{\Snorm{\hZ\cdot W^\varepsilon}}{\Vnorm{W^\varepsilon}}\to\infty.
\end{equation*}
\end{proof}
The connection between our definition of an ultra-relativistic vector
field and ${\bf v}\cdot{\bf v}\simeq 1$ is exhibited by introducing the
$1$-parameter family of Newtonian velocities ${\bf v}^\varepsilon$ of
$W^\varepsilon$ with respect to the observer field $U$:
\begin{align*}
{\bf v}^\varepsilon
= -\frac{W^\varepsilon}{W^\varepsilon\cdot U}-U
\end{align*}
where $U\cdot U=-1$.
It follows that
\begin{align*}
{\bf v}^\varepsilon\cdot{\bf v}^\varepsilon
=&
\left(\frac{W^\varepsilon}{W^\varepsilon\cdot U}+U\right)
\cdot
\left(\frac{W^\varepsilon}{W^\varepsilon\cdot U}+U\right)
=
\frac{W^\varepsilon\cdot W^\varepsilon}{(W^\varepsilon\cdot U)^2}
+
\frac{2(W^\varepsilon\cdot U)}{W^\varepsilon\cdot U}
+
U\cdot U
=
\frac{W^\varepsilon\cdot W^\varepsilon}{(W^\varepsilon\cdot U)^2}
+1.
\end{align*}
Thus $\lim_{\varepsilon\to0}{{\bf v}^\varepsilon}\cdot{{\bf v}^\varepsilon}=1\iff\lim_{\varepsilon\to0}{W^\varepsilon\cdot W^\varepsilon}/
{(W^\varepsilon\cdot
U)^2}=-\lim_{\varepsilon\to0}\left({\Vnorm{W^\varepsilon}/\Snorm{W^\varepsilon\cdot
U}}\right)^2 = 0
$ and hence $\lim_{\varepsilon\to0}{{\bf v}^\varepsilon}\cdot{{\bf v}^\varepsilon}=1\iff\lim_{\varepsilon\to0}\Snorm{W^\varepsilon\cdot
U}/{\Vnorm{W^\varepsilon}} \to \infty$ (i.e. $W^\varepsilon$ is
ultra-relativistic).

Observe that in definition (\ref{intro_def_ultra_rel}) $\lim_{\Vep\to
0}W^\Vep$ may or may not exist. 
Given a $1$-parameter vector field, $\calW^\varepsilon$,
$\calW^\varepsilon$ is said to be \defn{light-like limited} if
$\calW^0=\lim_{\varepsilon\to0} \calW^\varepsilon$ exists and is
light-like (i.e. $\calW^0\ne 0$
and $\Vnorm{\calW^0}=0$).
The relationship between ultra-relativistic vector
fields and light-like limited vector fields is
given by the following lemma:
\begin{lemma}
If $\calW^\varepsilon$ is light-like limited then $\calW^\varepsilon$
is ultra-relativistic. If $\calW^\varepsilon$ is ultra-relativistic
and the limit $\lim_{\varepsilon\to0} \calW^\varepsilon$ exists
and is nowhere $0$
then $\calW^\varepsilon$ is light-like limited.
\end{lemma}
\begin{proof}
If $\calW^\varepsilon$ is light-like limited then
$\lim_{\varepsilon\to0}\Snorm{\calW^\varepsilon\cdot Z}=
\Snorm{\lim_{\varepsilon\to0}(\calW^\varepsilon)\cdot Z}=
\Snorm{\calW^0\cdot Z}>0$ since $\calW^0$ is light-like and $Z$ is time-like.
Furthermore,
$\lim_{\varepsilon\to0}(\calW^\varepsilon\cdot\calW^\varepsilon)=-\Vnorm{\calW^0}^2=0$
and so $\calW^\varepsilon$ is ultra-relativistic.

If $\calW^\varepsilon$ is ultra-relativistic and the limit
$\calW^0=\lim_{\varepsilon\to0} \calW^\varepsilon$ exists and is
nowhere $0$
then (since $\calW^0\cdot Z$ is finite) $\Vnorm{\calW^0}$ must vanish
to ensure (\ref{intro_def_ultra_rel}) is obeyed.
\end{proof}

The $1$-parameter family of velocity fields $V^\Vep$ introduced in the main
body of this article (see section
\ref{section:construction_of_the_scheme}) is ultra-relativistic.
Moreover, the vector
fields $V_{-1}$, $V_0$ and $V_1$ then have certain properties:
\begin{lemma}
\label{lm_Vm1_ne_0}
$ $\\
\vspace{-1em}
\begin{itemize}
\item The $1$-parameter family of vector fields $V^\varepsilon$
is ultra-relativistic if and only if
$V_{-1}$ does not vanish anywhere.
\item If $V^\varepsilon$ is ultra-relativistic then $V_0$ is
nowhere time-like and $V_1$ does not vanish anywhere.
\end{itemize}
\end{lemma}
\begin{proof}
$ $\\
\vspace{-1em}
\begin{itemize}
\item
If $V_{-1}\ne 0$
then for any time-like vector $Z$, $V_{-1}\cdot Z\ne0$ so
\begin{align*}
\lim_{\varepsilon\to0}(\Snorm{V^\varepsilon\cdot Z}/\Vnorm{V^\varepsilon})=
\lim_{\varepsilon\to0}(\varepsilon^{-1} \Snorm{V_{-1}\cdot Z})=\infty
\end{align*}
since $\Vnorm{V^\varepsilon}=1$ and
so $V^\varepsilon$ is ultra-relativistic.
If there is some $x\in\man$
such that $V_{-1}(x)=0$
then
\begin{align*}
\lim_{\varepsilon\to0}(\Snorm{V^\varepsilon(x)\cdot Z(x)}/
\Vnorm{V^\varepsilon(x)})=
\lim_{\varepsilon\to0}(\varepsilon^{0} \Snorm{V_{0}(x)\cdot Z(x)})\ne\infty
\end{align*}
so $V^\varepsilon$ is not ultra-relativistic.
\item
Assume $V_0$ is time-like. Since $V^\varepsilon$ is ultra-relativistic
it follows $V_{-1}\ne 0$ and, furthermore, since $V_0$ is time-like
$V_{-1}\cdot V_0\ne0$; however, this conclusion contradicts
(\ref{np_lorentz_V_0}) which states $V_0$ and $V_{-1}$ are
orthogonal. Hence, $V_0$ is not time-like (i.e. $V_0$ is space-like,
light-like or zero).

Now since $V_{-1}$ is light-like and $V_{0}$ is not time-like then in
order to satisfy $2V_{-1}\cdot V_{1} + V_{0}\cdot V_{0} =-1$ from
(\ref{np_lorentz_V_1}) $V_{1}$ does
not vanish anywhere.
\end{itemize}
\end{proof}

\newcommand{\Div}{\bm{\nabla}\cdot}
\newcommand{\Curl}{\bm{\nabla}\times}
\newcommand{\eMKS}{{\bm{e}}}
\newcommand{\bMKS}{{\bm{b}}}
\newcommand{\Mom}{{\bm{p}}}
\newcommand{\eMKSvec}{{\mathbb E}}
\newcommand{\bMKSvec}{{\mathbb B}}
\newcommand{\Momvec}{{\mathbb P}}

\subsection{The perturbation hierarchy in Gibbs' $3$-vector notation}
\label{appendix:Gibbs_3-vector_version}
To facilitate a comparison with existing fluid descriptions of beam dynamics
\cite{davidson:1990, davidson:2001} that employ Gibbs' $3$-vector
notation the equations \Fieldsysintegrefs in section
\ref{section:construction_of_the_scheme} are here transcribed into
that language.

By splitting with respect to an inertial observer field
$\frac{1}{c}\PD_t$ where
\begin{equation*}
g = -c^2 dt\otimes dt + dx\otimes dx + dy\otimes dy + dz\otimes dz
\end{equation*}
one obtains
\begin{equation}
\begin{aligned}
&\Div\eMKS = \frac{m_0 c^2}{q_0}\,\gamma\rho,\\
&\Curl\bMKS=\frac{1}{q_0}\,\rho\Mom
+ \frac{1}{c^2}\frac{\PD\eMKS}{\PD t},\\
&\Curl\eMKS + \frac{\PD\bMKS}{\PD t} = 0,\\
&\Div\bMKS = 0,\\
&\gamma\frac{\PD\Mom}{\PD t} +
\left(\frac{\Mom}{m_0}\cdot\bm{\nabla}\right)\Mom =
q_0\left(\gamma\eMKS+\frac{1}{m_0}\Mom\times \bMKS\right),\\
&-\gamma^2+\frac{\Mom\cdot\Mom}{m_0^2 c^2} = -1,\\
&m_0\frac{\PD}{\PD t}(\gamma\rho)+\Div (\rho\Mom)=0
\end{aligned}
\label{MKS_Gibbs_system}
\end{equation}
and
\begin{equation}
\gamma\PD_t\gamma + \frac{\Mom}{m_0}\cdot\bm{\nabla}\gamma =
\frac{q_0}{m_0^2c^2}\Mom\cdot\eMKS
\label{MKS_Gibbs_energy_balance}
\end{equation}
where $\bm{a}\cdot\bm{b}$ is the standard
Euclidean dot product of two $3$-vector fields $\bm{a}$ and
$\bm{b}$. Note that (\ref{MKS_Gibbs_energy_balance}) can be obtained by taking the
scalar product of $\Mom$ with the differential equation for $\Mom$ and the
algebraic constraint on $\gamma$ and $\Mom$ in
(\ref{MKS_Gibbs_system}). Therefore, like the final equation in
(\ref{MKS_Gibbs_system}) (charge conservation), equation 
(\ref{MKS_Gibbs_energy_balance}) is
redundant. However, the two differential equations are treated very
differently in the subsequent asymptotic analysis. Equation
(\ref{MKS_Gibbs_energy_balance}) is
automatically maintained order-by-order by the algebraic constraint on
$\gamma$ and $\Mom$ but charge conservation must be imposed
order-by-order. Hence, (\ref{MKS_Gibbs_energy_balance}) is not
included in the following.

The scalar field $\rho$ is the reduced
proper charge density introduced in the main body of the article and
the MKS fields $(\Mom,\eMKS,\bMKS)$ and $\gamma$ are related to $(V,F)$ by
\begin{align*}
&V=\frac{\gamma}{c}\PD_t + \frac{\Momvec}{m_0 c}\\
&F=\frac{q_0}{m_0 c}dt\wedge\tilde\eMKSvec + \frac{q_0}{m_0}\star\left(dt\wedge\tilde\bMKSvec\right)
\end{align*}
where ($\Momvec, \eMKSvec, \bMKSvec)$ are the vector fields
\begin{align*}
&\Momvec = (\Mom\cdot\bm{i})\,\PD_x + (\Mom\cdot\bm{j})\,\PD_y +
(\Mom\cdot\bm{k})\,\PD_z,\\
&\eMKSvec = (\eMKS\cdot\bm{i})\,\PD_x + (\eMKS\cdot\bm{j})\,\PD_y +
(\eMKS\cdot\bm{k})\,\PD_z,\\
&\bMKSvec = (\bMKS\cdot\bm{i})\,\PD_x + (\bMKS\cdot\bm{j})\,\PD_y +
(\bMKS\cdot\bm{k})\,\PD_z
\end{align*}
and $(\bm{i},\bm{j},\bm{k})$ is an orthonormal inertial $3$-vector frame.

The expansions for the $1$-parameter families $\eMKS^\Vep$,
$\bMKS^\Vep$, $\gamma^\Vep$, $\Mom^\Vep$ and $\rho^\Vep$ corresponding
to (\ref{intro_V_R_F_expan}) are
\begin{align*}
&\eMKS^\Vep = \varepsilon^{-1}\eMKS_{-1} + \eMKS_{0} +
   \varepsilon\eMKS_{1} + \ldots\,,\qquad\
\bMKS^\Vep = \varepsilon^{-1}\bMKS_{-1} + \bMKS_{0} +
   \varepsilon\bMKS_{1} + \ldots\,,\qquad\
\\
&\gamma^\Vep = \varepsilon^{-1}\gamma_{-1} + \gamma_{0} +
   \varepsilon\gamma_{1} + \ldots\,,\qquad\
\Mom^\Vep = \varepsilon^{-1}\Mom_{-1} + \Mom_{0} +
   \varepsilon\Mom_{1} + \ldots\,,\quad\
\\
&\rho^\Vep = \varepsilon\rho_{1} + \varepsilon^2\rho_{2} + \ldots
\end{align*}
Substituting the above expansions into (\ref{MKS_Gibbs_system}) yields
the following hierarchy corresponding to equations
(\ref{np_maxwell_F_m1}-\ref{np_lorentz_V_1}):
\begin{steplist}
\item
The source-free Maxwell equations are satisfied by the
prescribed external electric field $\eMKS_{-1}$ and external magnetic
field $\bMKS_{-1}$:
\begin{align*}
&\Div\eMKS_{-1} = 0,\\
&\Curl\bMKS_{-1} =
\frac{1}{c^2}\frac{\PD\eMKS_{-1}}{\PD t},\\
&\Curl\eMKS_{-1} + \frac{\PD\bMKS_{-1}}{\PD t} = 0,\\
&\Div\bMKS_{-1} = 0.
\end{align*}
\item
The partial differential equation
\begin{align}
\gamma_{-1}\frac{\PD\Mom_{-1}}{\PD t} +
\left(\frac{\Mom_{-1}}{m_0}\cdot\bm{\nabla}\right)\Mom_{-1} =
q_0\left(\gamma_{-1}\eMKS_{-1}+\frac{1}{m_0}\Mom_{-1}\times \bMKS_{-1}\right),
\label{gibbs_step2_diff}
\end{align}
and algebraic condition
\begin{align}
-\gamma_{-1}^2+\frac{\Mom_{-1}\cdot\Mom_{-1}}{m_0^2 c^2} = 0
\label{gibbs_step2_alg}
\end{align}
are solved for $\gamma_{-1}$ and $\Mom_{-1}$.
\item
The leading order equation describing charge conservation is
\begin{align*}
m_0 \frac{\PD}{\PD t}(\gamma_{-1}\rho_{1}) + \Div(\rho_{1}\Mom_{-1}) =
0
\end{align*}
and is solved for $\rho_{1}$.
\item
Solve the Maxwell equations for $\eMKS_{0}$ and $\bMKS_{0}$ with
$\tfrac{m_0 c^2}{q_0}\,\gamma_{-1}\rho_{1}$ and
$\frac{1}{q_0}\rho_{1}\Mom_{-1}$ as sources:
\begin{align*}
&\Div\eMKS_{0} = \frac{m_0 c^2}{q_0}\,\gamma_{-1}\rho_{1},
\\
&\Curl\bMKS_{0}= \frac{1}{q_0}\,\rho_{1}\Mom_{-1}
+\frac{1}{c^2}\frac{\PD\eMKS_{0}}{\PD t},
\\
&\Curl\eMKS_{0} + \frac{\PD\bMKS_{0}}{\PD t} = 0,\\
&\Div\bMKS_{0} = 0.
\end{align*}

\item
Solve
\begin{align*}
&\gamma_{-1}\frac{\PD\Mom_{0}}{\PD t} +
\gamma_{0}\frac{\PD\Mom_{-1}}{\PD t} +
\left(\frac{\Mom_{0}}{m_0}\cdot\bm{\nabla}\right)\Mom_{-1} +
\left(\frac{\Mom_{-1}}{m_0}\cdot\bm{\nabla}\right)\Mom_{0}
\\
&\qquad\qquad\qquad\qquad
=
q_0\left(\gamma_{-1}\eMKS_{0}+\gamma_{0}\eMKS_{-1}\right)+
\frac{q_0}{m_0}\left(\Mom_{-1}\times\bMKS_{0}+
\Mom_{0}\times\bMKS_{-1}\right),
\\
&-\gamma_{-1}\gamma_{0}+
\frac{\Mom_{-1}\cdot\Mom_{0}}{m^2_0 c^2} = 0
\end{align*}
for $\gamma_{0}$ and $\Mom_{0}$. Use of the above
algebraic equation to eliminate $\gamma_0$ leads to an inhomogeneous
\emph{linear} partial differential equation for $\Mom_0$.
\item
The equation
\begin{align*}
m_0 \frac{\PD}{\PD t}(\gamma_{0}\rho_{1})+
m_0 \frac{\PD}{\PD t}(\gamma_{-1}\rho_{2}) + \Div(\rho_{1}\Mom_{0})
+\Div(\rho_{2}\Mom_{-1})
=0
\end{align*}
enforcing charge conservation is solved for $\rho_{2}$.
\item
The electric and magnetic fields $\eMKS_{1}$ and $\bMKS_{1}$ are
solutions to the following Maxwell equations:
\begin{align*}
&\Div\eMKS_{1} = \frac{m_0 c^2}{q_0}
\left(\gamma_{-1}\rho_{2}+\gamma_{0}\rho_{1}\right),
\\
&\Curl\bMKS_{1}=
\frac{1}{q_0}\left(\rho_{2}\Mom_{-1}+\rho_{1}\Mom_{0}\right)
+
\frac{1}{c^2}\frac{\PD\eMKS_{1}}{\PD t},
\\
&\Curl\eMKS_{1} + \frac{\PD\bMKS_{1}}{\PD t} = 0,\\
&\Div\bMKS_{1} = 0.
\end{align*}

\item
Solve
\begin{align*}
&\gamma_{-1}\frac{\PD\Mom_{1}}{\PD t} +
\gamma_{0}\frac{\PD\Mom_{0}}{\PD t} +
\gamma_{1}\frac{\PD\Mom_{-1}}{\PD t} +
\left(\frac{\Mom_{-1}}{m_0}\cdot\bm{\nabla}\right)\Mom_{1} +
\left(\frac{\Mom_{0}}{m_0}\cdot\bm{\nabla}\right)\Mom_{0} +
\left(\frac{\Mom_{1}}{m_0}\cdot\bm{\nabla}\right)\Mom_{-1}
\\&\qquad\qquad\qquad\qquad=
q_0\left(\gamma_{-1}\eMKS_{1}+\gamma_{0}\eMKS_{0}+\gamma_{1}\eMKS_{-1}\right)+
\frac{q_0}{m_0}\left(\Mom_{-1}\times\bMKS_{1}+\Mom_{0}\times\bMKS_{0}+
\Mom_{1}\times\bMKS_{-1}\right),
\\
&-2\gamma_{-1}\gamma_{1}-\gamma_{0}\gamma_{0}+
\frac{2\Mom_{-1}\cdot\Mom_{1}+\Mom_{0}\cdot\Mom_{0}}{m^2_0 c^2} = -1
\end{align*}
for $\gamma_{1}$ and $\Mom_{1}$. Eliminating $\gamma_1$ using the
above algebraic equation leads to an inhomogeneous \emph{linear} partial
differential equation for $\Mom_{1}$.
\end{steplist}

\end{appendix}

\end{document}